\documentclass[aps,pra,showpacs,twoside,twocolumn,10pt,superscriptaddress,nofootinbib,reprint,floatfix,%
amsmath,amssymb,%
floatfix,
]{revtex4-2}
\usepackage{hyperref}
\usepackage[T1]{fontenc} 
\usepackage{nicefrac}
\usepackage{graphicx}
\usepackage{dcolumn}
\usepackage{bm}
\usepackage{xcolor}
\usepackage[ruled,vlined,linesnumbered]{algorithm2e}
\usepackage{setspace}
\usepackage[noend]{algpseudocode}
\usepackage{svg}

\newcommand{\bra}[1]{\ensuremath{\left\langle#1\right|}}
\newcommand{\ket}[1]{\ensuremath{\left|#1\right\rangle}}

\begin{document}

\title{Improving success probability in the LHZ parity embedding by computing with quantum walks.}
\author{Jemma Bennett} 
\email{jemma.bennett@uibk.ac.at}
\affiliation{Institute for Theoretical Physics, University of Innsbruck, Technikerstra{\ss}e 21A, 6020 Innsbruck, Austria}
\author{Nicholas Chancellor}
\affiliation{School of Computing, Newcastle University, 1 Science Square, Newcastle upon Tyne, NE4 5TG, UK}
\author{Viv Kendon}
\affiliation{Department of Physics, University of Strathclyde, Glasgow, G4 0NG, UK}
\author{Wolfgang Lechner}
\affiliation{Institute for Theoretical Physics, University of Innsbruck, Technikerstra{\ss}e 21A, 6020 Innsbruck, Austria}
\affiliation{Parity Quantum Computing GmbH, Rennweg 1/Top 314, 6020 Innsbruck, Austria}

\begin{abstract}

The LHZ parity embedding is one of the front-running methods for implementing difficult-to-engineer long-range interactions in quantum optimisation problems. 
Continuous-time quantum walks are a leading approach for solving quantum optimisation problems. Due to them populating excited states, quantum walks can avoid the exponential gap closing problems seen in other continuous-time techniques such as quantum annealing and adiabatic quantum computation (AQC).
An important question therefore, is how continuous-time quantum walks perform in combination with the LHZ parity embedding.
By numerically simulating continuous-time quantum walks on 4, 5 and 6 logical qubit Sherrington-Kirkpatrick (SK) Ising spin glass instances embedded onto the LHZ parity architecture, we are able to verify the continued efficacy of heuristics used to estimate the optimal hopping rate and the numerical agreement with the theory behind the location of the lower bound of the LHZ parity constraint strength. 
In addition, by comparing several different LHZ-based decoding methods, we were able to identify post-readout error correction techniques which were able to improve the success probability of the quantum walk.

\end{abstract}

\maketitle

\section{Introduction}

In the quantum optimization setting, we aim to solve useful optimization problems by mapping them to all-to-all connected Ising models. 
However, beyond a very small number of qubits, in current hardware this level of connectivity does not exist.
For example on the Zephyr topology of the current DWave machines, the largest complete graph that it is possible to directly embed is $n=4$ qubits \cite{Boothby2021}.
In comparison, IBM machines use the `heavy-hex' hardware graph, which favours low connectivity in order to reduce errors caused by frequency collisions and cross-talk. Here the largest complete graph that it is possible to directly embed is $n=2$ \cite{Chamberland2020}.
In order to overcome the issues caused by this lack of connectivity in hardware, we have to use an embedding.
Several methods have been proposed. The approach favoured on DWave machines is minor embedding \cite{choi08a, choi10a, minorminer}. This involves connecting chains of qubits, such that several physical qubits represent a single one in the optimization problem. However, as the problem size scales, larger chain sizes are required. A larger chain means the qubits within it are more susceptible to flipping, and therefore increasing chain strength is required, which will eventually lead to scaling problems. 
Another option is to use perturbative gadgets \cite{Mozgunov2023} to enact interactions, however, the strength of links between qubits for this technique does not scale favourably.

Alternatively, embeddings which convert long-range couplings into locally connected structures have been proposed.
In \cite{Palacios2024}, configurations of three qubit triangles within an Ising model are encoded onto pairs of coupled qubits, which are then connected and constrained such that the low energy eigenspace of the original model is preserved.
Another local encoding is the LHZ parity embedding \cite{Lechner2015}, which is the focus of the research in this paper. Here, the parity of each of the couplings in an Ising model is represented by a single qubit. The strength of the coupling is enacted by a field on the qubit. Three or four body constraints are used to suppress states which don't correspond to states in the original Ising model, i.e., physical states which don't lie inside the logical codespace.
Due to these constraints, this embedding has some protection against error, as long as the strength of the constraints is strong enough that the eigenenergies of the physical states which lie outside the logical codespace are well separated from those which are within the codespace \cite{Lanthaler2021}.
However, the high required strength of constraints leads to scalability concerns, as well as the possibility of a single bit-flip error cascading across several qubits. 

The redundancy gained from using the LHZ embedding may also be used for error mitigation or correction. Several different decoding methods may be used. In \cite{Weidinger2023, Weidinger2024}, spanning trees were used to perform error mitigation for QAOA on the LHZ architecture. Spanning trees were previously suggested and utilised for error mitigation in the continuous-time setting in \cite{Lechner2015, Albash2016}.
When spanning trees are used to decode from a physical to logical state, each tree measured represents one possible `choice' of logical state.
The next question, is how to use this set of choices to select the solution.
Prior to this selection, there is also the opportunity to use the set of choices as candidate solutions which are stored and then eventually utilized, as was described in \cite{King2019}.
In situations where just the energy (and not the solution state) is required, we may take the mean energy of the choices as the solution \cite{Weidinger2024}.
Where just one choice is selected, a decision must be made on how to select this choice.
In most of the previous implementations, the final logical state was selected via a majority vote on the `choices' given by the measured spanning trees. 
This approach is advantageous when the error rate is low and less excited states are populated (as you would expect in AQC), and you are more likely to measure a majority of correct states than not. Conversely, if the error rate is high, and more excited states are populated, an approach where you find or approximate the lowest energy state may be more beneficial. 
In \cite{Weidinger2024}, a method is presented where the spanning tree with the minimum energy is selected as the solution.
For the research in this paper, we compared the performance of decoding the spanning trees using the majority vote method versus the lowest energy chosen method.

In \cite{Albash2016, Pastawski2016, Messinger2024, Nambu2024a, Nambu2024b, Nambu2025}, the performance of several belief-propagation-type decoders for error correction on the LHZ embedding were investigated. In \cite{Albash2016}, minimum weight and maximum likelihood decoders were also investigated.
Where the logical state must be preserved, for example during a continuing computation, a sequence of CNOT gates, or a combination of X-measurements and Z-rotations may be used to preserve and act upon parity information \cite{Fellner2022, Messinger2023, Smith2024}.

Quantum walks (QW) \cite{Farhi1998, Childs2004} are a form of continuous-time quantum computing (CTQC), which also includes adiabatic quantum computing (AQC) and quantum annealing.
In general, all types of CTQC use a Hamiltonian of the form,
\begin{equation}
\hat{H}(t) = A(t)\hat{H}_0 + B(t)\hat{H}_P,
\end{equation}
where $\hat{H}_0$ is an easily implemented initial Hamiltonian and $\hat{H}_P$ is the problem Hamiltonian which has the solution to an optimization problem encoded into its ground state. 
$\hat{H}_0$ defines the graph of how the possible states are connected. 
There are multiple choices of $\hat{H}_0$ but in this paper we use,
\begin{equation}
    \hat{H}_0 = n\mathbb{I} - \sum_{j=0}^{n-1}\hat{X}_j,
\end{equation}
which defines the $n-$dimensional hypercube where $n$ is the number of qubits.  
In AQC and quantum annealing the functions $A(t)$ and $B(t)$ are time-dependent and varied from 1 to 0 and 0 to 1 respectively. By contrast, for quantum walks both $A(t)$ and $B(t)$ are time-independent and set to $A(t)=\gamma$ and $B(t)=1$, where $\gamma$ is known as the ``hopping rate'' and determines the rate at which the quantum walk can pass through the graph of possible states. A more comprehensive introduction to optimisation using continuous-time quantum walks may be found in \cite{Callison2019, Morley2017}.
In \cite{Callison2019}, it was shown that quantum walks can do better than Grover's search algorithm when finding the ground states of Sherrington-Kirkpatrick spin glasses. More recently multistage quantum walks have been shown to improve performance of quantum walks and outperform QAOA in some cases \cite{Callison2020, Banks2023, Schulz2023, Gerblich2024}.

In \cite{Ghosh2024}, it was suggested that the non-adiabaticity in quantum walks leads to possible superpositions of  correct states, which  can avoid the effects of first order phase transitions caused by the closing of the gap in adiabatic techniques.
Results in \cite{Bennett2023} suggest that continuous-time quantum walks may have some intrinsic robustness to error due to low-lying excited states populated during the quantum walk. These states have the potential to be used in error mitigation techniques. 
Furthermore, in \cite{Nambu2024a, Nambu2024b, Nambu2025}, a combination of a Monte Carlo method \cite{Nambu2022} and belief-propagation and approximation of belief-propagation (bit-flipping) decoding algorithm (using syndromes defined in \cite{Pastawski2016}) were used to find logical ground states of LHZ parity embedded spin glass instances. When using the belief-propagation step of their decoding technique, they found that a substantial number of the physical states they were able to decode to the logical ground state lay outside the logical codespace.
In addition, they found the optimal constraint strength when using the belief-propagation based decoding method was different to when it was not used. These results may indicate that different decoding methods may be better suited to different error conditions.

In this paper we performed numerical simulations of continuous-time quantum walks on 4, 5, and 6 logical qubit SK spin glasses (that were first generated and studied in \cite{Callison2019, Bennett2023b}) embedded onto the LHZ parity architecture. Using our simulations, we were able to evaluate the efficacy of the heuristic method (developed in \cite{Callison2020}) for estimating the optimal hopping rate of the quantum walk  in the LHZ setting. Additionally we numerically tested the theory on the lower bound of constraint strength in the LHZ parity embedding (developed in \cite{Lanthaler2021}) in the quantum walk setting.
Utilising the redundancy present in the LHZ parity embedding, we compared the performance on the success probability of several pre-existing post-readout error correcting decoding methods: entire state decoding, random overlapping spanning trees, non-overlapping spanning trees, minimum weight decoding and belief propagation. 
In addition, following the research in \cite{Weidinger2024}, we utilised the idea of decoding using lowest energy rather than majority votes, for decoding methods which give multiple choices of logical state. We then applied this innovation to the spanning tree decoding methods and compared the performance.
As a sanity check, we compared the performance of all methods to the ideal but experimentally unrealistic directly embedded SK Ising spin glass models.

The paper is organised as follows. In section \ref{sec:Q_walks} we introduce our SK Ising spin glass optimization problem setting and how to solve it using a continuous-time quantum walk. 
In section \ref{sec:gamma-heur} we introduce and describe the calculation of the heuristic hopping rate $\gamma_{\text{heur}}$.
In section \ref{sec:LHZ_embedding} we introduce and describe the LHZ parity embedding and how it is applied in our problem setting. 
In section \ref{sec:const_strength} we analyse the effect of the constraint strength from the LHZ embedding on the performance of the quantum walk.
In section \ref{sec:decoding_methods} we introduce and describe the different decoding methods that we investigated in this research. 
In section \ref{sec:decode_perform} we analysed the performance of the different decoding methods on four different instances of 5 qubit SK Ising spin glasses. 
In section \ref{sec:prob_vs_size} we show how the performance of the different decoding methods varies with problem size. 
Finally in section \ref{sec:conc_and_future} we make conclusions and suggestions of future directions.

\section{Ising model}\label{sec:Q_walks}

\begin{figure}[!tb]
    \centering
    \includegraphics[width=0.3\textwidth]{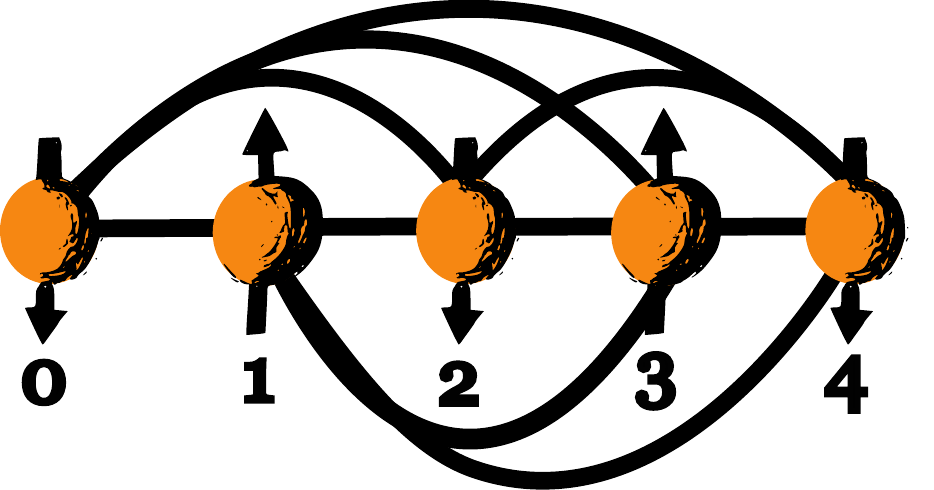}
    \caption{Diagram of a 5 qubit Sherrington-Kirkpatrick Ising spin glass}
    \label{fig:5qb_spin_glass}
\end{figure}

The solution to an optimization problem may be encoded into the ground state of an Ising model. In order to simulate finding the solution to difficult optimization problems, we use a data-set of Sherrington-Kirkpatrick (SK) spin glasses. Using the original (direct) embedding, they have a Hamiltonian of the form,
\begin{equation}
         \hat{H}_{P} = -\frac{1}{2}\sum_{j\neq k=0}^{n-1}J_{jk}\hat{Z}_j\hat{Z}_k -
         \sum_{j=0}^{n-1}h_j\hat{Z}_j.
\end{equation}
We then embedded the same instances according to the LHZ parity architecture and compared the performances.

 Figure \ref{fig:5qb_spin_glass} shows a 5 qubit all-to-all connected SK spin glass of the directly embedded form. The black lines represent couplings between the qubits (in orange) and the the qubits are labelled from 0 to 4. The data-set used in this paper was first generated for research reported in \cite{Callison2019} and can be found in a data archive \cite{Chancellor2019a_data}.
As a quantum walk is non-adiabatic, we expect in general that the probability of finding the correct ground state of $\hat{H}_p$ (the success probability) will be less than one. Therefore we need to take repeats to improve the success probability. As the success probability varies over the course of the quantum walk, we measure it at multiple times and take the average, which we call the average success probability $\bar{P}$. If we measure the probability over the interval $[t, t+\Delta t]$, we can define the average success probability as, 
\begin{equation}\label{eq:avg_succ_prob}
    \bar{P}(t, \Delta t) \equiv \frac{1}{\Delta t} \int_t^{t+\Delta t} dt_f P(t_f),
\end{equation}
where $t_f$ is the final time at which we measure the probability.

\begin{figure}
    \centering
    \includegraphics[width=0.43\textwidth]{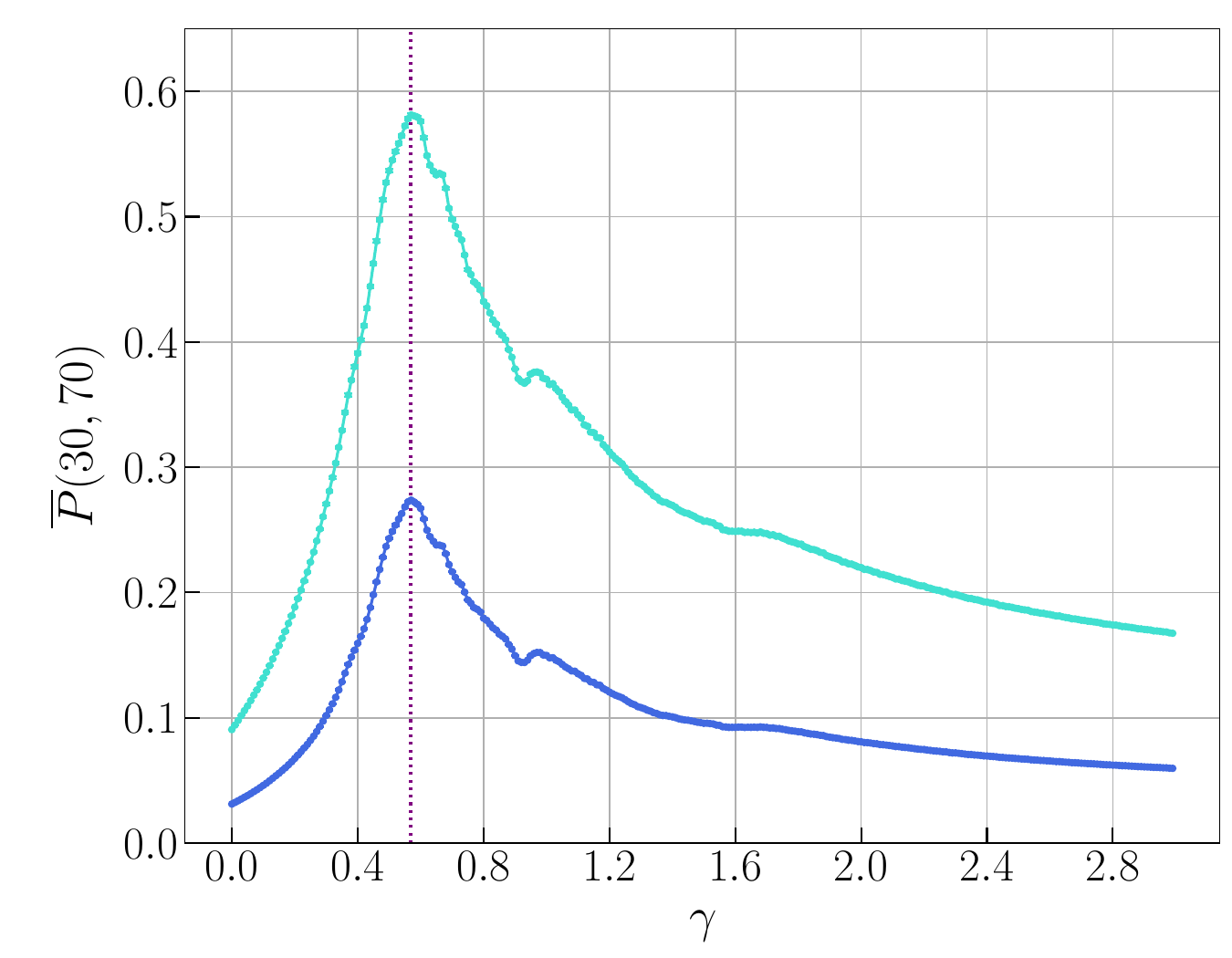}
    \caption{Graph showing the average success probability $\bar{P}(t, \Delta t)$ where $t=30$ and $\Delta t=70$ vs the hopping rate $\gamma$. For one copy of a 5 qubit spin glass instance with uid ``aagtzdpchgtknrzilrnhpvxqtvqiql'' in dark blue and for three separate copies of the same instance in light blue. Error bars for $\bar{P}(t, \Delta t)$ are plotted but are too small to be seen. The purple vertical dotted line indicates the location of the optimal hopping rate $\gamma_{\text{opt}}$ where $\bar{P}(t, \Delta t)$ is maximised.}
    \label{fig:plong_vs_gamma}
\end{figure}

An important variable in quantum walks is the hopping rate $\gamma$. This controls the rate at which the quantum walk passes through $\hat{H}_0$ the hypercube of possible states.  Figure \ref{fig:plong_vs_gamma} shows $\bar{P}(t, \Delta t)$ where $t=30$ and $\Delta t=70$ vs $\gamma$ for a single 5 qubit spin glass instance plotted in dark blue. To embed the 5 qubit spin glass onto the LHZ architecture, 15 qubits are required (see figure \ref{fig:5qb_spin_glass_LHZ}). This is an equivalent number of qubits to three separate copies of the directly embedded 5 qubit spin glass. Therefore, we also plot $\bar{P}(t, \Delta t)$ vs gamma for the three separate copies of the 5 qubit Ising model instance in light blue. A purple vertical dotted line indicates the location of the optimal hopping rate $\gamma_{\text{opt}}$ where $\bar{P}(t, \Delta t)$ is maximised.

\subsection{Heuristic $\gamma$}\label{sec:gamma-heur}

It can be seen from figure \ref{fig:plong_vs_gamma}, that $\bar{P}$ is highly dependent on $\gamma$. 
For the Ising spin glass problem, the exact value of $\gamma_{\text{opt}}$ cannot be calculated efficiently. However in \cite{Callison2020} a heuristic method for estimating optimal hopping rate $\gamma_{\chi}$ was developed.

There, they introduce the ``dynamic coefficient'' $\chi^{jk}$, which is a measure of the dynamics experienced by a system. It can be defined as,
\begin{equation}\label{eq:single_chi}
    \chi^{(jk)} = \frac{\zeta_{jk}/\Gamma(t)}{[1 + \zeta_{jk}/\Gamma(t)]^2},
\end{equation}
where $\zeta_{jk} = \frac{|\Delta_{jk}|}{2|\bra{k}\hat{H}_{0}\ket{j}|} $ is a single scaled gap parameter, with $|\Delta_{jk}|$ the energy difference between states $j$ and $k$.
As $\hat{H}_0$ is not biased then $|\bra{k}\hat{H}_{0}\ket{j}| = 1$ when $j, k$ differ by one bit-flip.
If we average across all of the pairs of states $j, k$ in a system we get the average dynamic coefficient, 
\begin{equation}\label{eq:avg_chi}
\bar{\chi} = \langle \chi^{(jk)} \rangle_{jk}.
\end{equation}
Due to the exponentially growing number of states, it is not efficient to calculate \eqref{eq:avg_chi}. However, we are able to approximate $\bar{\chi}$ by talking a sample of the available states $j, k$. The error in this approximation is,
\begin{equation}\label{eq:err_chi}
    \delta \bar{\chi} \sim \frac{0.25}{N^{1/2}_{\text{samples}}},
\end{equation}
where $N_{\text{samples}}$ is the number of samples.

\begin{figure}
    \centering
    \includegraphics[width=0.43\textwidth]{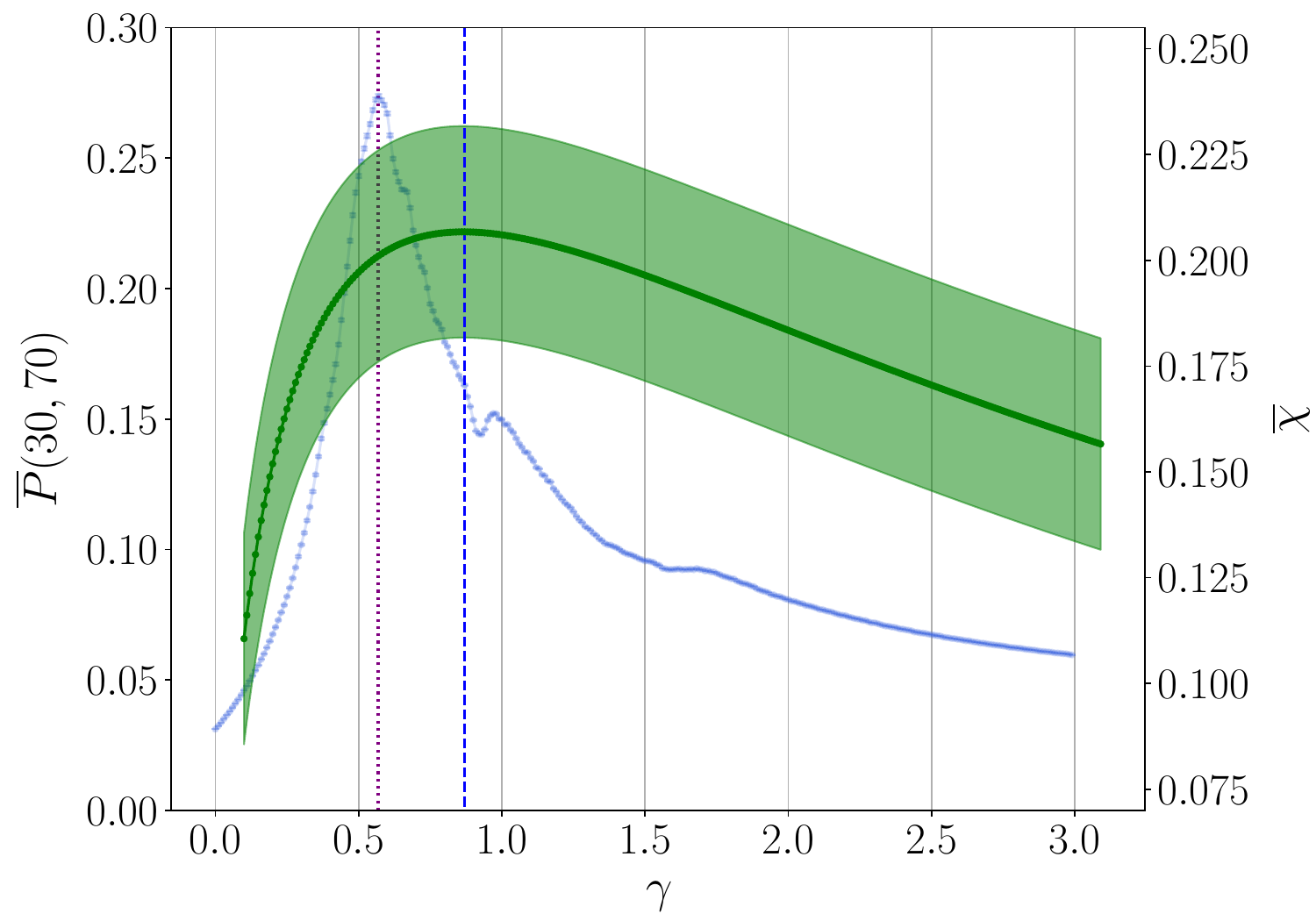}
    \caption{ Graph showing the average dynamic coefficient $\bar{\chi}$ (on the right hand y axis vs the hopping rate $\gamma$ for a single directly embedded 5 qubit spin glass instance with uid ``aagtzdpchgtknrzilrnhpvxqtvqiql'' in green. Light green shading indicates the error in $\bar{\chi}$ calculated using equation \ref{eq:avg_chi}. The blue vertical dashed line indicates the location of the heuristic hopping rate $\gamma_{\text{heur}}$ where $\bar{\chi}$ is maximised. For comparison $\bar{P}(30, 70)$ (on the left hand y axis) vs $\gamma$ is also plotted in faded blue, as well as a purple vertical dotted line which indicates the location of $\gamma_{\text{opt}}$.}
    \label{fig:dyns_vs_gamma}
\end{figure}

We can estimate a heuristic value for the optimal hopping rate $\gamma_{\text{heur}}$ by finding $\gamma$ which maximises $\bar{\chi}$. Figure \ref{fig:dyns_vs_gamma} shows $\bar{\chi}$ (on the left hand y axis) of a single 5 qubit spin glass instance for values of $\gamma$ between 0 and 3.2, plotted in dark green. Light green shading indicates the error in $\bar{\chi}$ calculated using \eqref{eq:err_chi}, where $N_{\text{samples}}=10^3$. A blue dotted vertical line indicates the value of $\gamma$ at which $\bar{\chi}$ is maximised. For comparison we have also plotted $\bar{P}(t, \Delta t)$, where $t=30$ and $\Delta t=70$, vs $\gamma$ for the same 5 qubit directly embedded instance (same data as figure \ref{fig:plong_vs_gamma}). A purple vertical dotted line indicates the location of $\gamma_{\text{opt}}$. 

From figure \ref{fig:dyns_vs_gamma}, we can see that the value of $\gamma_{\text{heur}}$ does not directly coincide with $\gamma_{\text{opt}}$. However, due to the width of the peak in $\bar{P}(t, \Delta t)$, for this 5 qubit spin glass instance, when using $\gamma_{\text{heur}}$ as the hopping rate of the quantum walk, there will still be some improvement in the average success probability $\bar{P}$.

\section{LHZ parity architecture}\label{sec:LHZ_embedding}

\begin{figure}[!tb]
    \centering
    \includegraphics[width=0.45\textwidth]{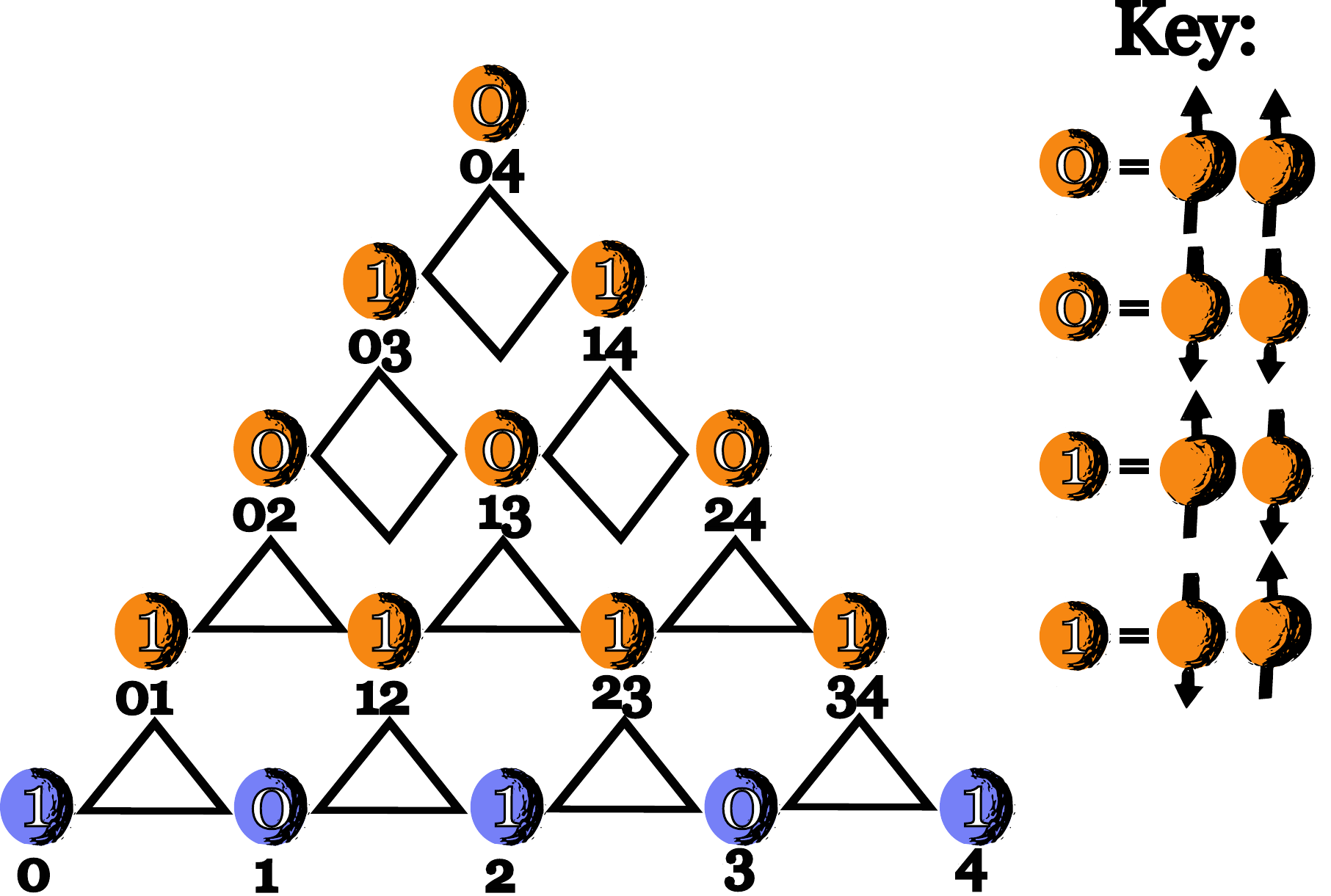}
    \caption{Left: Diagram of a 5 logical qubit Sherrington-Kirkpatrick Ising spin glass embedded onto the LHZ architecture. The numbers below each of the orange qubits refer to which coupling in the original they represent. The numbers below the blue qubits represent the qubit on which each of the fields act. Squares (and triangles) represent 4 (and 3) body constraints. Right: Key showing the parity conversion of couplings to LHZ qubits.}
    \label{fig:5qb_spin_glass_LHZ}
\end{figure}

The LHZ parity embedding was developed as a way to embed Ising models with all-to-all connectivity onto a locally connected graph.
In order to carry out the LHZ parity embedding, we start by mapping each of the $J_{ij}\hat{\sigma}^z_i\hat{\sigma}^z_j$ couplings between qubits $i$ and $j$ in the $n$ qubit logical Ising model onto single `coupling' qubits with a local field of the form $J_{ij}\tilde{\sigma}^z_{ij}$. Each of the new qubits representing a coupling in the original Ising model can be seen in on the diagram in the left of figure \ref{fig:5qb_spin_glass_LHZ}, labelled as $ij$.
Each of the coupling qubits is given a spin of either 0 or 1 depending on the parity of the qubit spins in the original Ising model. The right-hand diagram in figure \ref{fig:5qb_spin_glass_LHZ} shows this relationship.
If both spins in the original (direct) embedding were the same, the coupling qubit is given a spin of 0, but if the two spins were opposite the coupling qubit is given a spin of 1.

The coupling qubits are arranged in a triangular structure (the LHZ triangle) in a specific order around triangles or squares. These triangles and squares, represent loops of three or four couplings in the original Ising model and the parity constraints in the LHZ embedding.
The parity constraints are added to suppress states which are now possible in the parity embedding (due to the increased number of qubits) but were impossible in the original Ising model. 
For example, if we look at the top square of the LHZ triangle in figure \ref{fig:5qb_spin_glass_LHZ}, in this model, (without the constraints) the state where qubits \textbf{03, 13, 14, 04} have the values \textbf{1, 0, 1, 1} respectively, would be possible. However, on the original Ising model, having this sequence of parities in a loop of 4 is impossible.

Following the above reasoning, you can convince yourself that an odd number of spin 1 qubits in a loop of three or four in the parity embedding results in an impossible state in the original Ising model. Therefore constraints of the form, $C(\Tilde{\sigma}^z_{(l, 1)}\Tilde{\sigma}^z_{(l, 2)}\Tilde{\sigma}^z_{(l, 3)}(\Tilde{\sigma}^z_{(l, 4)}))$ are introduced, with the fourth $\tilde{\sigma}^z$ depending on whether it is a loop of three or four couplings. A constraint will become equal to $+C(1)$ if an odd number of ones are present in a state, thereby penalising its existence. Thus these states are suppressed.

For the final step of the parity embedding, we add the local field qubits $h_i\hat{\sigma}_i$ from the original direct embedding of the Ising model to the LHZ parity embedding. As can be seen in figure \ref{fig:5qb_spin_glass_LHZ} (in blue), we do this by simply mapping $h_i\hat{\sigma}_i \rightarrow h_i\tilde{\sigma}_i$.
We then connect these local field (data) qubits to the coupling qubits via a triangle constraint. The coupling qubit directly above each two local field qubits corresponds to the coupling between them in the original Ising model. This ensures the parity between the two data qubits is consistent with that on the coupling qubit.

\subsection{Constraint strength}\label{sec:const_strength}

If we wish to fully suppress the unwanted states using the constraint terms, we must ensure that the lower energy levels of the physical LHZ Hamiltonian match those of the logical problem Hamiltonian. There is an minimal value of constraint strength such that these states are suppressed while minimising the strength of the penalty (and energy scale of the system). A method to determine this minimal value was described in \cite{Lanthaler2021}.

\begin{figure}
    \centering
    \includegraphics[width=0.46\textwidth]{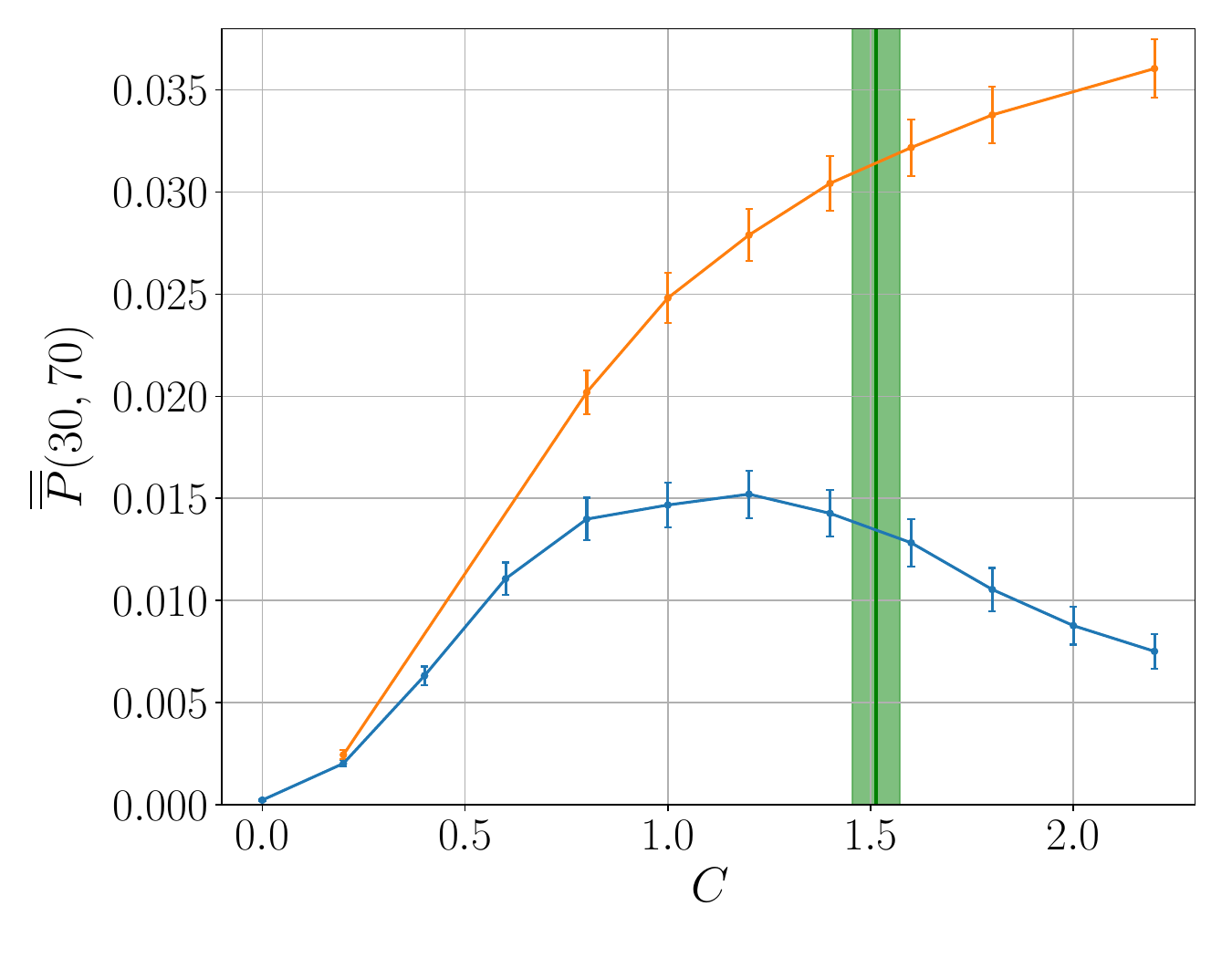}
    \caption{Graph showing $\overline{P}(30, 70)$ at $\gamma_{\text{heur}}$ (blue) and at $\gamma_{\text{opt}}$ (orange) averaged over 100 instances of 5 logical qubit SK Ising spin glasses embedded onto the LHZ parity architecture versus the constraint strength $C$, which was varied from 0.0 to 2.2. The instances were decoded using the entire state decoding method. Error bars on each data point indicate the uncertainty in $\overline{\overline{P}}(30, 70)$. Also plotted (green vertical line) is the lower bound of $C$ averaged across the instances calculated using the method in \cite{Lanthaler2021}. Green shading indicates the error in this value.}
    \label{fig:probs_vs_constraint}
\end{figure}

In order to get an idea of how the constraint strength should be set for our problem set in the quantum walk setting, we next measured the performance of the quantum walk whilst varying the constraint strength, on a set of 5 logical (15 physical) qubit SK Ising spin glasses. Figure \ref{fig:probs_vs_constraint} shows the average long-time success probability of the quantum walk averaged over 100 instances $\overline{\overline{P}}(30, 70)$ of 5 logical qubit SK Ising spin glasses embedded onto the LHZ parity architecture versus the constraint strength $C$, which was varied from 0.0 to 2.2.
Plotted in blue, $\overline{\overline{P}}(30, 70)$ was calculated using the heuristic estimate of optimal hopping strength, $\gamma_{\text{heur}}$ (calculated using the method described in section \ref{sec:gamma-heur} from \cite{Callison2020}). 
For comparison, plotted in orange, $\overline{\overline{P}}(30, 70)$  was calculated using the optimal value of hopping strength $\gamma_{\text{opt}}$. 
Also plotted (green, vertical line), is the average value of the lower bound of $C$ (calculated using the method in \cite{Lanthaler2021}) across the 100 instances. The green shading indicates the standard error.

We might naively expect the value of constraint strength which provides the highest average success probability for the quantum walk $\overline{\overline{P}}(30, 70)$ to fall above the average calculated value of the lower bound of the constraint strength $C$. Indeed, looking at figure \ref{fig:probs_vs_constraint}, for $\overline{\overline{P}}(30, 70)$ calculated using $\gamma_{\text{opt}}$, this is the case. However,  we see that for $\overline{\overline{P}}(30, 70)$ calculated using $\gamma_{\text{heur}}$, the optimal constraint strength for the quantum walk falls below the average lower bound. 

\begin{figure}
    \centering
    \includegraphics[width=0.45\textwidth]{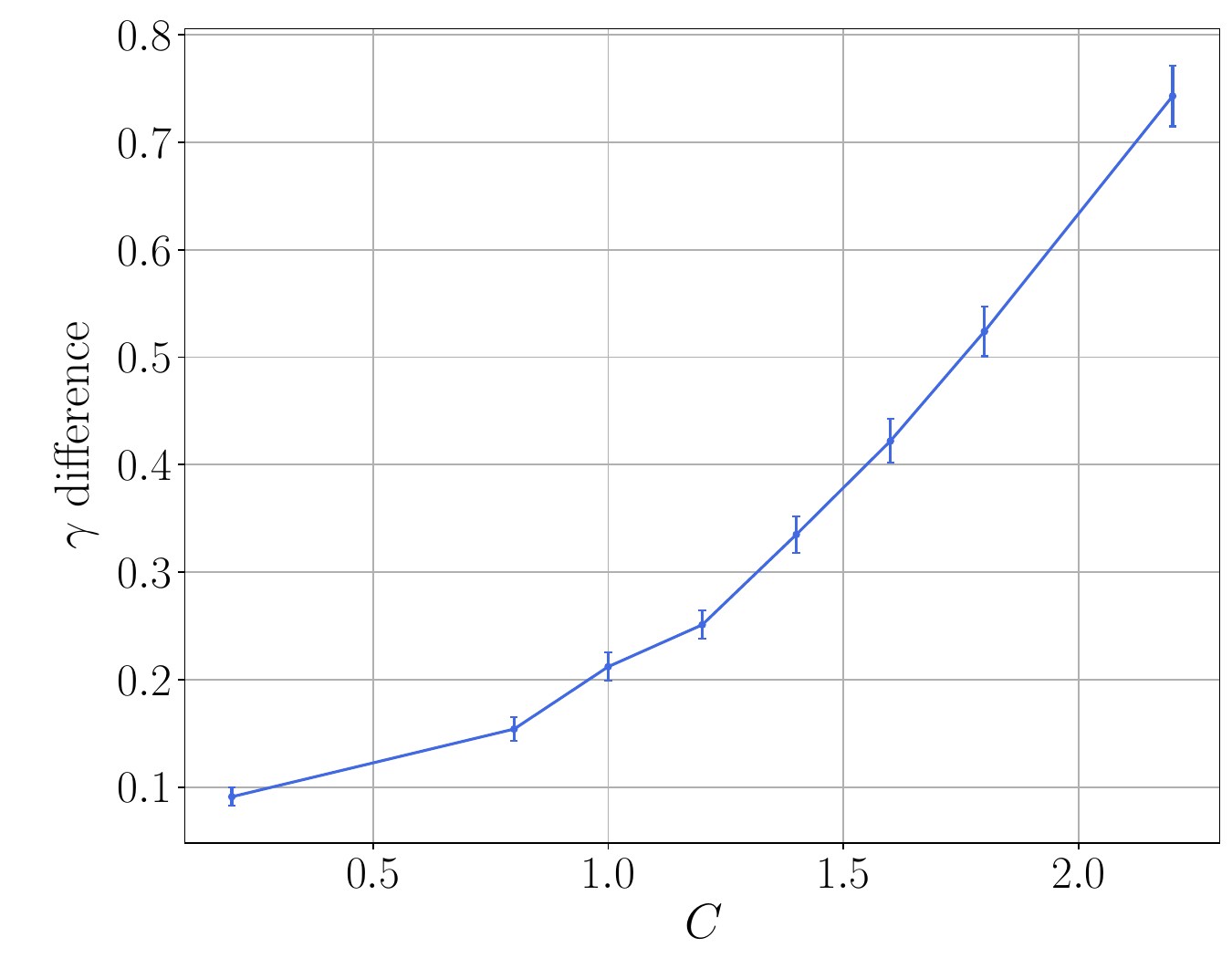}
    \caption{Graph showing the average difference between $\gamma_{\text{opt}}$ and $\gamma_{\text{heur}}$ ($\gamma$ difference) versus the constraint strength for 100 instances of 5 logical qubit LHZ embedded SK Ising spin glasses. Error bars on each point indicate the uncertainty in the $\gamma$ difference. }
    \label{fig:gamma_diff_vs_constraint}
\end{figure}

Looking at this counter-intuitive finding, we note that the success probability of the quantum walk is also dependent on the proximity of the the heuristic hopping rate $\gamma_{\text{heur}}$ to the optimal hopping rate $\gamma_{\text{opt}}$ ($\gamma$ difference). In order to verify whether this was the cause of the discrepancy seen here, we next measured the average difference between $\gamma_{\text{heur}}$ and $\gamma_{\text{opt}}$ across the same 100 instances of 5 logical qubit LHZ embedded SK Ising spin glasses, for values of constraint strength between 0.2 and 2.2. Results are shown in figure \ref{fig:gamma_diff_vs_constraint}.

Looking at this figure, we see that in agreement with our expectations the $\gamma$ difference increases with increasing constraint strength. In addition, we see that the difference increases with an accelerating rate. We argue that this trend is the cause of the discrepancy between the optimal constraint strength when $\overline{\overline{P}}(30, 70)$ is measured using $\gamma_{\text{heur}}$ and the calculated average lower bound.
This is further backed up by the fact that we do not see this feature when $\overline{\overline{P}}(30, 70)$ is measured using $\gamma_{\text{opt}}$.

\section{Decoding}\label{sec:decoding_methods}

Once the Ising spin glass (the logical problem) has been embedded onto the LHZ parity architecture, the quantum computation may then be carried out, in order to find its physical ground state. In this paper we carry out this computation using a continuous-time quantum walk as described in section \ref{sec:Q_walks}.  
Once the physical ground state of the LHZ system has been computed, a decoding step must then be performed, in order to recover a logical state of the Ising spin glass. This state should correspond to the ground state of the directly embedded SK Ising spin glass. There are several options of how to decode the LHZ embedded physical state to a logical one. In this section, we discuss several of the possible options. 

\subsection{Entire state decoding}

One option is to compare the entire physical ground state that has been computed, with the physical ground state that corresponds exactly to the ground state of the directly embedded logical SK Ising spin glass. If the entire physical ground state doesn't correspond to the logical ground state, then we discard it as incorrect. This is most intuitive option as it is simply the reverse of embedding the logical problem onto the LHZ parity architecture. However, this method is sensitive to errors, as no constraints may be violated. Even just one bit-flip can cause the state to be incorrect. 
This approach is problematic, when a technique such as a quantum walk is used for computation. As a quantum walk populates some excited states when it is carried out, it is likely that a final measured state may contain some bit-flips compared to the entirely correct state. This decoding method also does not make use of the redundancy available in the LHZ embedding.

\subsection{Random overlapping spanning trees}

\begin{figure}[!tb]
    \centering
    \includegraphics[width=0.43\textwidth]{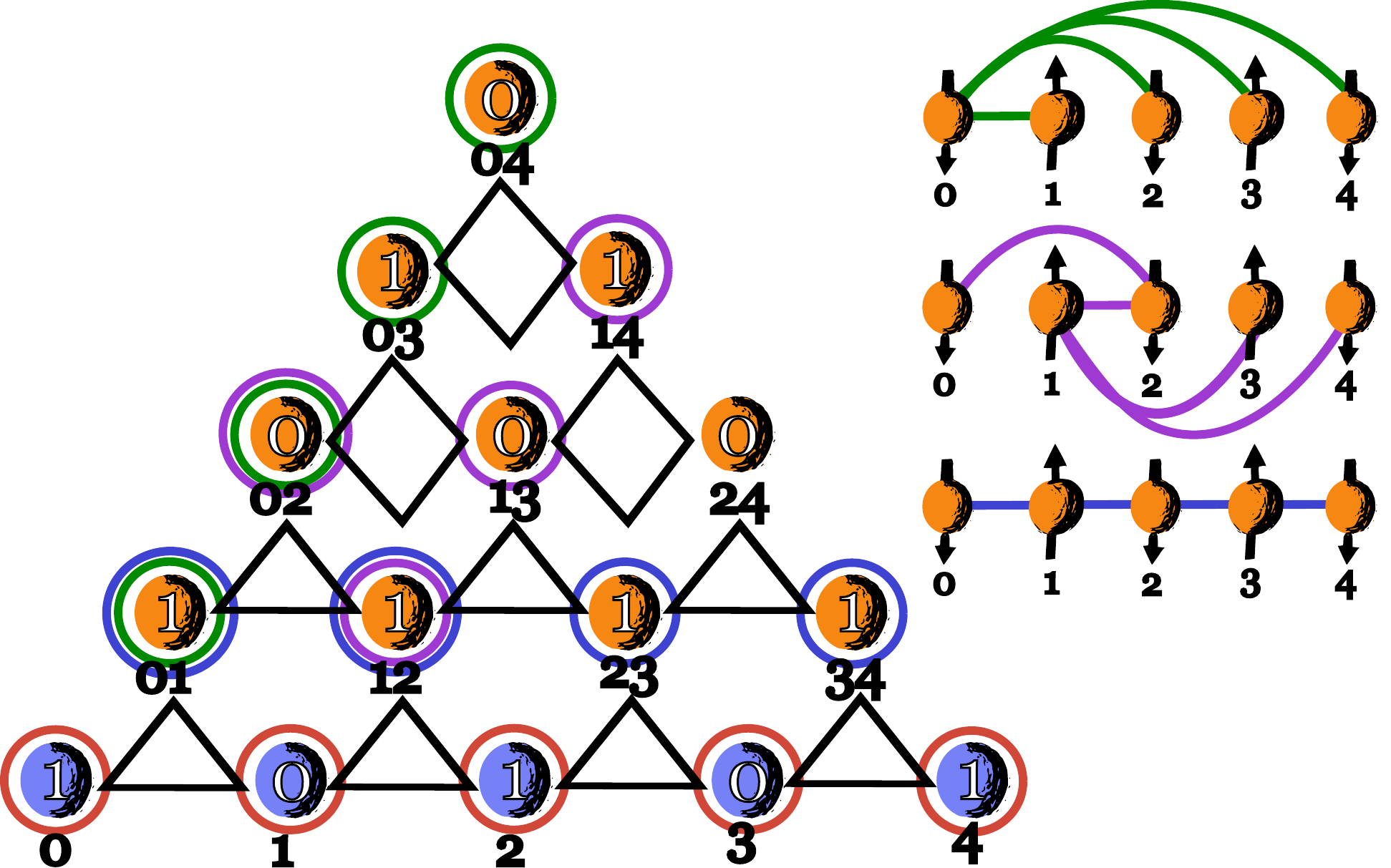}
    
    \caption{Left: Diagram of the LHZ parity embedded 5 logical qubit SK Ising spin glass, with three possible choices of overlapping spanning trees. The qubits of each of the three trees are circled in green, purple and blue. Also, the data qubits are circled in red to highlight their use as one `choice'. Right: Diagram highlighting the links of the respective green, purple and blue spanning trees as they would appear on the original directly embedded Ising spin glass.}
    \label{fig:5qb_spanning_tree_LHZ}
\end{figure}

Another option is to consider the state of one or more parts of the physical ground state which correspond to `spanning trees' in the logical Ising spin glass graph. 
A spanning tree is a subset of a graph where each node is connected to at least one other node and there are no cycles. 
The values of the physical qubits in the LHZ system which make up a spanning tree may be used to define a possible ground state of the logical Ising spin glass (up to a global bit/spin flip).
Unlike entire state decoding, this decoding method allows us to still define a possible ground state even when constraints in the LHZ system have been violated, as is likely using a quantum walk.
Also we are able to define multiple spanning trees per LHZ triangle, thereby utilising the redundancy of the embedding.

When decoding using spanning trees, there are multiple options of how to choose the trees. One option is to choose the trees randomly and allow them to overlap. Overlapping means that a physical qubit participates in more than one spanning tree.
The top right of figure \ref{fig:5qb_spanning_tree_LHZ} shows three possibilities of overlapping spanning tree, in green, purple and blue, on the directly embedded 5 qubit spin glass.  Each of the couplings in each spanning tree on the directly embedded model correspond to qubits (circled in the same colour) in the LHZ embedding, shown on the left hand side of the figure.
Each of the blue data qubits is also circled in red, indicating their use as an extra `choice' of the possible ground state of the original directly embedded model, which is included in our implementation of this decoding method.

We see in figure \ref{fig:5qb_spanning_tree_LHZ} that qubits 01, 02 and 12 have two rings of more than one colour surrounding them. This indicates that they participate in two overlapping spanning trees. 
If the spanning trees are chosen randomly, overlapping qubits become more likely as the number of spanning trees increases and are inevitable after the number of spanning trees exceeds $K/(n-1)$, where $K$ is the number of physical qubits and $n$ is the number of logical qubits.
If the spanning trees are allowed to be overlapping, then the number of possible spanning trees are limited to Cayley's number ($n^{(n-2)}$). 

For most of the research carried out in this paper, the decoding method we call `random overlapping spanning trees' consists of selecting two random spanning trees (which are allowed to overlap) and then also measuring the values of the data qubits as a third `choice'. We denote this as `2+1' random overlapping spanning trees.
Where we investigated the effect of an increasing number of randomly selected spanning trees which were allowed to overlap (see appendix \ref{sec:success_vs_n_trees}), we continued to carry out a  single measurement of the data qubits which counted once toward the total number of trees.

Once the spanning trees have been measured, they must then be `decoded' such that there remains only one single option as the computed logical ground state. In previous research \cite{Lechner2015, Albash2016, Weidinger2023}, a majority vote on the spanning tree states was performed in order to confirm which was the chosen ground state. In \cite{Weidinger2023}, the spanning tree states were also used to inform future steps of a QAOA approach. In \cite{Weidinger2024}, the performance of decoding spanning trees using their mean or lowest energy was also investigated. In this paper we compare the performance of a `majority votes' to a `lowest energy chosen' decoding step. 

We also note the classical cost of this decoding method. Each time a further spanning tree is measured a classical cost is levied. 
It is therefore important for the sake of fair testing that this classical cost is properly accounted for in comparisons. In appendix \ref{apx:spannning_cost} we provide analysis of the classical cost of measuring an increasing number of spanning trees. Then in appendix \ref{sec:find_sp_trees} we define a semi-analytical method of counting the total number of possible states introduced by measuring all the spanning trees.
In appendix \ref{sec:success_vs_n_trees} we investigated the effect on the quantum walk success probability of decoding using an increasing number of spanning trees and compared this to the success probability when decoding a randomly chosen state using the same number of spanning trees, for 100 instances of 4 and 5 logical qubit SK spin glasses.

\subsection{Non-overlapping spanning trees}

\begin{figure}[!tb]
    \centering
    \includegraphics[width=0.45\textwidth]{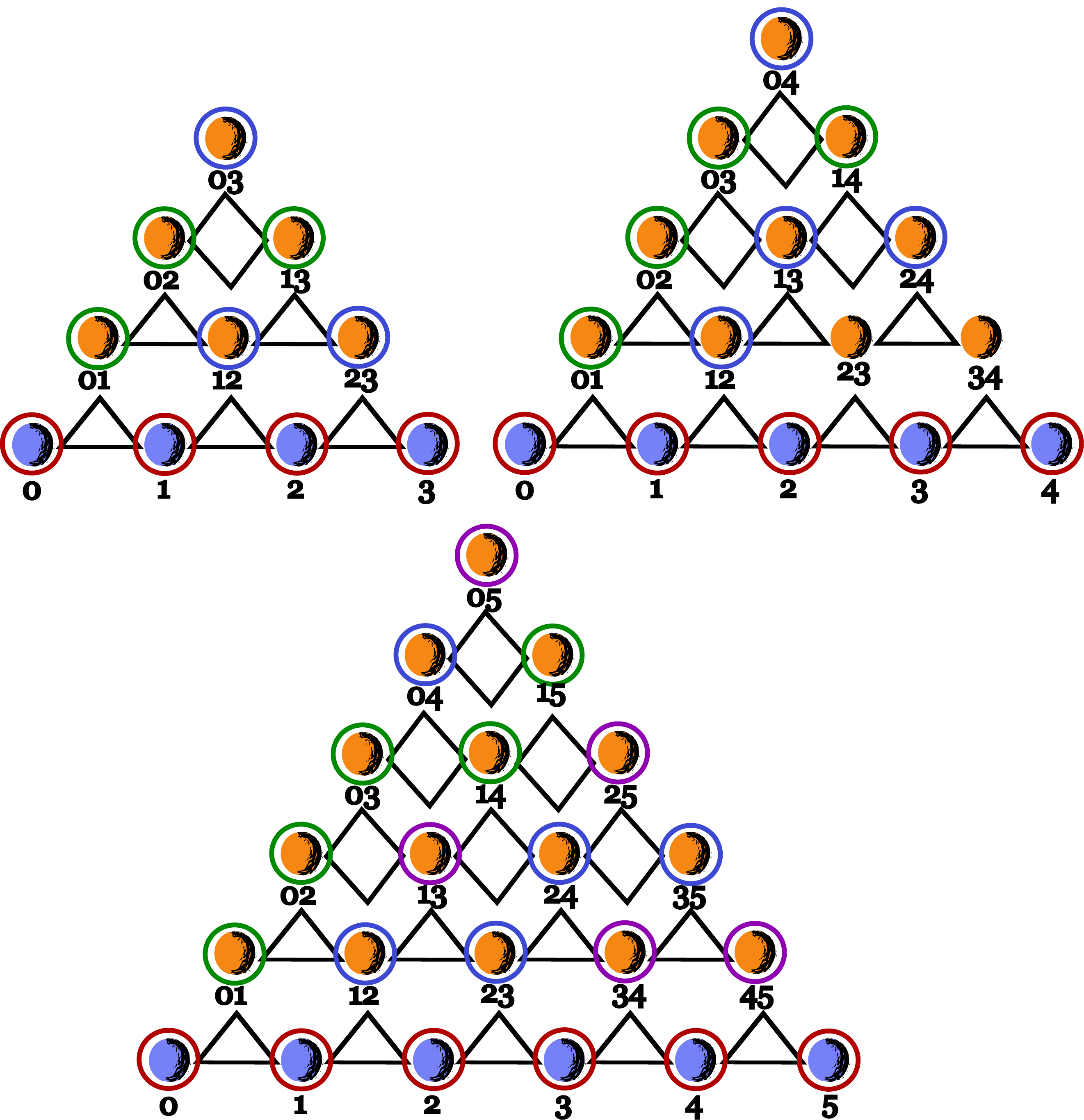}
    
    \caption{Diagram of possible choices of readout qubits (in red, green, blue and purple) for non-overlapping spanning trees for 4, 5, and 6 logical qubit SK spin glass problems embedded onto LHZ triangles.}
    \label{fig:non_overlap_trees}
\end{figure}

Another issue with using random overlapping spanning trees to decode is that with increasing number of spanning trees the chances of overlapping qubits increases, thereby reducing the efficiency of the method.
It is also not clear how to limit the number of spanning trees measured as the problem size increases. (The total number of spanning trees grows with $n^{(n-2)}$ so it is not feasible to measure them all at each size.)

One way to avoid these issues is to use non-overlapping spanning trees.
Here, instead of defining the spanning trees randomly, we define the spanning trees methodically such that there are no `overlapping' qubits, avoiding the inefficiency. For a LHZ embedding with $K$ physical qubits, the number of possible non-overlapping spanning trees is $\lfloor (K/(n-1)\rfloor$, where n is the number of logical qubits. This limits the number of possible spanning trees to a more feasible value.

Figure \ref{fig:non_overlap_trees} shows two (4 and 5 logical qubits) and three (6 logical qubits) possible choices of non-overlapping spanning trees chosen methodically. Again, we represent the spanning trees by surrounding the each of physical qubits (in the LHZ embedding) with circles of a specific colour (in this case: blue, green and purple). 
Like for the random overlapping spanning trees, we have encircled the blue data qubits of each of the LHZ embedded examples with red rings. This indicates that we again used the values of the data qubits as a `choice' of the possible ground state of the model.

To compare the performance of the non-overlapping spanning trees fairly with the random overlapping spanning trees, we also fixed our method to measuring two spanning trees and the values of the data qubits as a third `choice'. We denote this as `2+1' non-overlapping spanning trees.
We note that there are multiple ways to define non-overlapping spanning trees, so the choices shown in figure \ref{fig:non_overlap_trees} are not unique. 

In order to define the non-overlapping spanning trees used in this research, we developed the following method.
For this method, we consider the LHZ triangle without the data qubits.
First we find the total number of non-overlapping spanning trees to be defined: $s= \lfloor K/(n-1)\rfloor$. We then look at the `logical lines' of the LHZ triangle. These are the `lines' within the LHZ triangle which contain the same index \cite{Rocchetto2016, Fellner2022}. 
Next, starting with logical line zero, we select qubits where their secondary index (i.e. the 1 in 01) does not yet appear in any of the secondary indices of the previously selected qubits, until, in this logical line, we are left with a number of qubits equal to the remaining number of spanning trees to be defined (after this one). 
We then move to the next logical line along and repeat this process until all possible secondary indices have appeared. If the set of qubits we have now selected corresponds to a spanning tree, we go back to the start of the method and start defining the next spanning tree.
Conversely, if the set of qubits we have now selected does not correspond to a spanning tree, ensuring that we do not overfill a logical line, we select the qubit/s with the lowest index that will complete the spanning tree. We then go back to the start of this method and continue to define spanning trees until we have defined the total number of non-overlapping spanning trees $s$.
In the research described here, we used two of the non-overlapping spanning trees defined in this way at each problem size for our `2+1' non-overlapping spanning trees decoding method. 

Like for the random overlapping spanning trees, once the spanning trees had been measured, there was a choice of how to decode their output. Just as for the random overlapping spanning trees, we compared the performance of `majority votes' and `lowest energy chosen' decoding methods.

\subsection{Minimum weight decoder}

Another possible approach to decoding the ground state of the LHZ embedded Ising spin glasses is minimum weight decoding. Here, the physical state is measured and then a correction suggested which minimises the number of bit flips in order to return a physical state outside the logical codespace to the logical codespace. 

We can write this approach as its own optimization problem. If the number of plaquettes in the LHZ embedded model is given by $P = \frac{(N-1)(N-2)}{2} + (N-1)$, then we can loop through each of the possible $P =2^{P} - 1$ incorrect syndromes $s^c$ (i.e. when at least one of the plaquettes ${\sigma}^z_{(l, 1)}\Tilde{\sigma}^z_{(l, 2)}\Tilde{\sigma}^z_{(l, 3)}(\Tilde{\sigma}^z_{(l, 4)})) = -1$) in the set $S$, finding the minimum number of bit flip correction/s for each syndrome.  

To set up the Hamiltonian $H_{MW}$ which has the minimum bit flip correction to the physical state, we first set all the local fields to one. This way all qubits prioritise being in the zero state. 
Then we connect these qubits with the same four-body terms as in the original problem, but with each of the terms constraint strength set to its corresponding value in the current syndrome $s^c$ multiplied by a constant $\lambda$ which ensures that the plaquette values will all be +1 in its ground state. We can then write this Hamiltonian as,
\begin{equation}\label{eq:Hmw}
    H_{MW} = -\sum_{j=0}^{N-1}\Tilde{\sigma}^z_j - \sum_{v=0}^{P-1} s^c_v \lambda(\Tilde{\sigma}^z_{(v, 1)}\Tilde{\sigma}^z_{(v, 2)}\Tilde{\sigma}^z_{(v, 3)}(\Tilde{\sigma}^z_{(v, 4)})).
\end{equation}

We first note that the solution to $H_{MW}$ is likely to be degenerate, i.e. a single syndrome has multiple correction options. Therefore in this research we pick one random correction for each syndrome. Looking at the performance of the minimum weight decoder with multiple solutions could be an interesting avenue for future research.

In \cite{Albash2016}, equation \eqref{eq:Hmw} was shown to be equivalent to MAX-2-SAT and therefore NP-hard, which means its computational intensity will grow exponentially with problem size. On the other hand, since this method is problem instance independent, it is perhaps viable for problems where the same size LHZ parity embedding will be used multiple times.

\subsection{Belief propagation decoder}

The final decoding technique researched here is the belief propagation decoder which was first described and researched (for the purposes of LHZ decoding) in \cite{Pastawski2016}.
Here, instead of using the plaquette constraints, separate three parity qubit constraints are defined which themselves imply the values of three logical qubits. This is repeated such that the value of each logical qubit is implied by three of these constraints.

In \cite{Pastawski2016} it was found that this form of decoding had good results under independent and identically distributed noise. However, it was consequently shown in \cite{Albash2016}, using a simulated annealing quantum Monte Carlo method that this was not a realistic noise model for quantum annealing.
Nevertheless, this decoding technique has had success in when decoding LHZ for gate-based applications \cite{Messinger2024}. More recently, investigations using the belief decoder and an adjusted form of it with a rejection free Markov chain Monte Carlo method for simulated annealing, have shown that it can be effective especially under non-optimal values of parity constraints \cite{Nambu2024a, Nambu2024b, Nambu2025}.

\section{Performance comparison of decoding techniques}\label{sec:decode_perform}

\begin{figure}
    \centering
\includegraphics[width=0.45\textwidth]{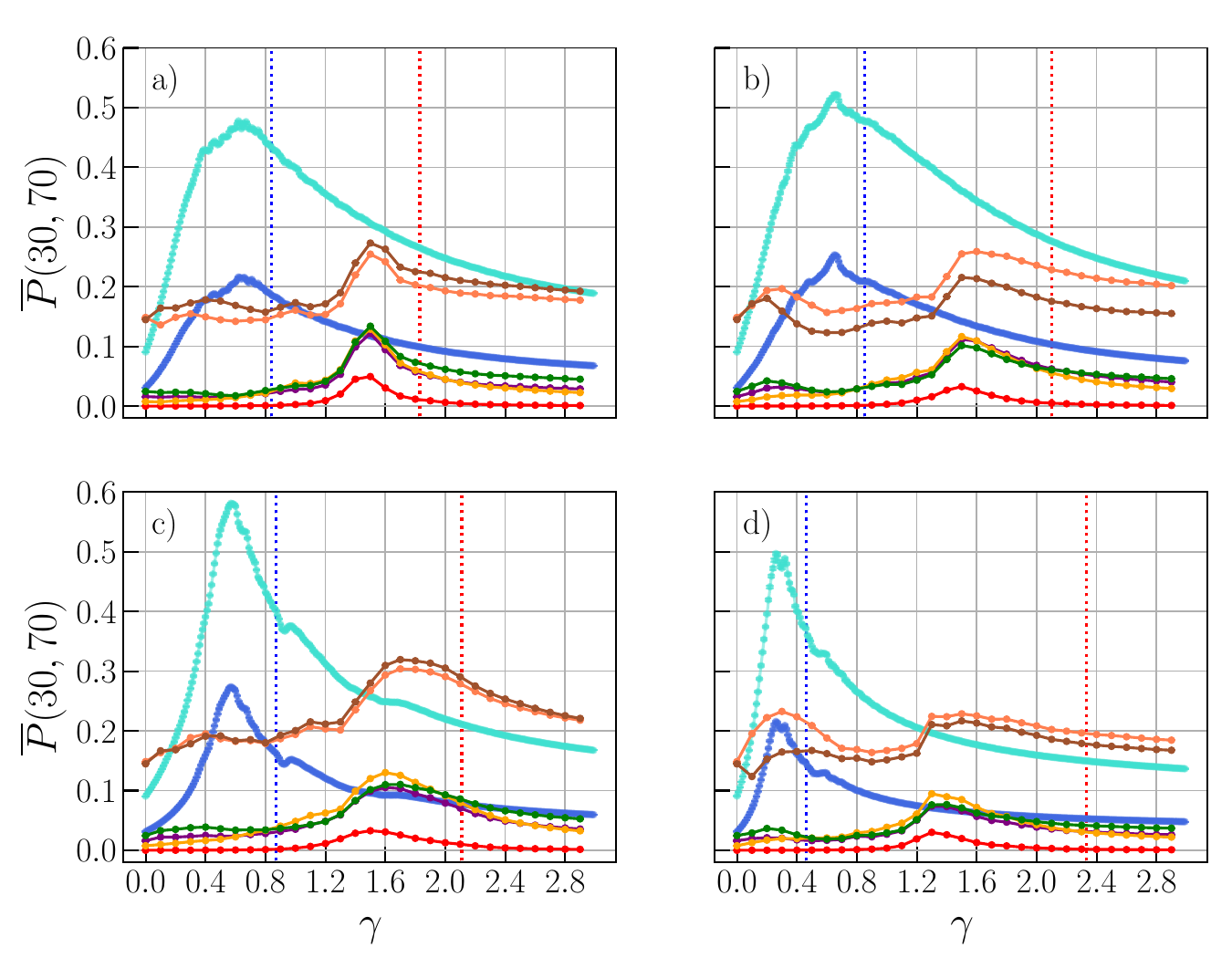}
    \caption{ Graph showing the success probability $\bar{P}(30, 70)$ versus $\gamma$ for four different instances of LHZ embedded 5 qubit SK spin glasses decoded (from physical to logical state) in six ways: entire state decoding (red), three random (overlapping) spanning trees (brown), three non-overlapping spanning trees (salmon), three non-overlapping spanning trees with majority vote (gold), the belief decoder (purple) and minimum weight decoder (green). For comparison $\bar{P}(30, 70)$ vs $\gamma$ for the same instances but with direct embedding are plotted: single copy (dark blue), three copies (light blue). 
    The $\gamma_{\text{heur}}$ for the directly embedded instances (dotted vertical blue) and LHZ embedded instances (dotted vertical red) are also plotted. These were calculated by maximising the average dynamic coefficient $\bar{\chi}$ as described in section \ref{sec:gamma-heur}.}
    \label{fig:LHZ_plong_vs_gamma}
\end{figure}

We next investigated how the performance of the computations using continuous-time quantum walk on four different instances of LHZ parity embedded 5 qubit SK spin glasses decoded with six different methods compared to their directly embedded counterparts. 
The four 5 qubit SK spin glass instances were first generated for research in \cite{Callison2019}  and can be found in data archive \cite{Chancellor2019a_data}, where they may be identified using their uids: `aaavmaiqiolnplcovmzxjazkyvyayz', `aacrpcjsbugeteaageltzcpnpovkcm', `aagtzdpchgtknrzilrnhpvxqtvqiql', `aakxejqunlcpqhmnftnrckailrczyp'. We carried out the parity embedding as explained in section \ref{sec:LHZ_embedding}, setting the constraint strength to $C=2$.

Figure \ref{fig:LHZ_plong_vs_gamma} a)-d) shows the long-time success probability $\bar{P}(t, \Delta t)$ (calculated using \eqref{eq:avg_succ_prob}), where $t=30$ and $\Delta t=70$, versus the hopping rate $\gamma$  for four instances of 5 qubit SK spin glass embedded onto the LHZ parity architecture. We compared the performance of six decoding methods: entire state decoding (red), $2+1$ random overlapping spanning trees with lowest energy chosen (brown), $2+1$ non-overlapping spanning trees with lowest energy chosen (salmon), $2+1$ non-overlapping spanning trees with majority vote (gold), a minimum weight decoder (green) and a belief decoder (purple).
For comparison the $\bar{P}(30, 70)$ vs $\gamma$ was also plotted for one (dark blue) and three (light blue) copies of the corresponding directly embedded instances.

The average dynamic coefficient for each of the directly and LHZ embedded instances was calculated using the method outlined in section \ref{sec:gamma-heur} and used to find the heuristic optimal hopping rate for the directly embedded instances $\gamma_{\text{heur}}^{\text{direct}}$ and the heuristic optimal hopping rate for the LHZ embedded instances $\gamma^{\text{LHZ}}_{\text{heur}}$. These were plotted as dotted vertical lines in blue and red respectively.
We notice that the location of $\gamma^{\text{LHZ}}_{\text{heur}}$ is increased compared to $\gamma_{\text{heur}}^{\text{direct}}$. This is not unexpected as we use the difference in energy between states $|\Delta_{jk}|$  in our calculation of $\chi^{jk}$, from which the heuristic optimal hopping rate is taken. The constraint terms in the LHZ embedding introduce larger values of $|\Delta_{jk}|$, thereby increasing the value of $\gamma^{\text{LHZ}}_{\text{heur}}$.

We also observe that the values of the true optimal hopping rate for the LHZ embedded instances $\gamma_{\text{opt}}^{\text{LHZ}}$ are increased compared to the the directly embedded instances $\gamma_{\text{opt}}^{\text{direct}}$ of the directly embedded models. This may be because the value of the true optimal hopping rate is known to be a balance in the the energy-scales of the transverse field $\hat{H}_0$ and problem Hamiltonian $\hat{H}_P$.
The constraint terms introduced by the LHZ embedding increase the energy scale of the problem Hamiltonian, thereby shifting the point of balance. 
As the heuristically calculated values of optimal hopping rate also show an increase for the LHZ embedding as well as remaining close to the true optimal hopping rate, this indicates that the heuristic method is still functioning satisfactorily.

For the LHZ embedded instance decoded using entire state decoding, we note that the success probability $\bar{P}(t, \Delta t)$ across all $\gamma$ is the lowest of all the decoding methods and is much reduced compared to the single directly embedded instance. This is not unexpected as the increased number of qubits involved in the LHZ embedding (15 compared to 5) increases the number of possible states in the system. But, as we require the entire state to be correct, there is still only one `correct' ground state, which the quantum walk must find. As when computing by continuous-time quantum walk, many of the excited `incorrect' states are also populated, we would expect the population in the correct ground state to be reduced, as is seen here. 

For the LHZ embedded instance decoded using the minimum weight decoder, the belief propagation decoder or the `2+1' non-overlapping spanning trees with majority vote, the success probability shows a slight improvement compared to the entire state decoder. This is because these decoders are able to decode physical states with unsatisfied constraints (which would register as incorrect to the entire state decoder) into logical states which may correspond to the correct ground state, thereby improving the success probability.

On the other hand for both spanning tree with lowest energy chosen decoding methods, we see the most improvement compared to the entire state decoding. We hypothesise this is the case for the following reasons. For the entire state decoder, from a possible $2^{15} = 32768$ states, there are only $2^5 = 32$ states which correspond to logical states. This means if a state is chosen randomly there is a $\frac{32}{32768} = 9.8\times 10^{-4}$ chance of it being a logical state, and a $\frac{1}{32768} = 3.1\times 10^{-5}$ chance of it being the correct logical state. Compare this with the directly embedded model where there is a $\frac{1}{32}=0.031$ chance of randomly finding the correct correct ground state.  

With the minimum weight decoder, belief decoder and the `2+1' spanning trees with majority vote, we aim to decode one physical state into one logical state. If there are $2^{15}=32768$ physical states, then, if the physical states are all decoded equally toward each possible logical state, there should be $32768/32 = 1024$ physical states per logical state. 
$\frac{1024}{32768}$ is equivalent to $\frac{1}{32}$ meaning there should be a 0.031 chance of finding the correct ground state randomly (same as the directly embedded model). In reality, as the decoder may not decode states evenly to each logical state or be able to decode every physical state to a logical one, this probability will be reduced. This is likely why we see the reduction in probability in figures \ref{fig:LHZ_plong_vs_gamma} a)--d) compared to the single directly embedded instance.

Alternatively, for the `2+1' random and non-overlapping spanning trees with lowest energy chosen, the aim is to decode a physical state into three choices of logical state.
If each of the choices are different, then one physical state has the possibility of decoding into three different logical states.
Although not all physical states will decode into three different logical states, those that do, have increased their chance of success. This suggests that physical states that have three different different logical state choices are still improving the success probability when just one of their choices corresponds to the correct logical ground state. This theory is supported by the fact that in figure \ref{fig:LHZ_plong_vs_gamma} we see that if we use a majority vote to decode the `2+1' non-overlapping spanning tree decoder, we lose the improvement in success probability compared to choosing with the lowest energy.

\subsection{Success probability by number of spanning trees correct}

\begin{figure}
    \centering
    \includegraphics[width=0.56\textwidth]{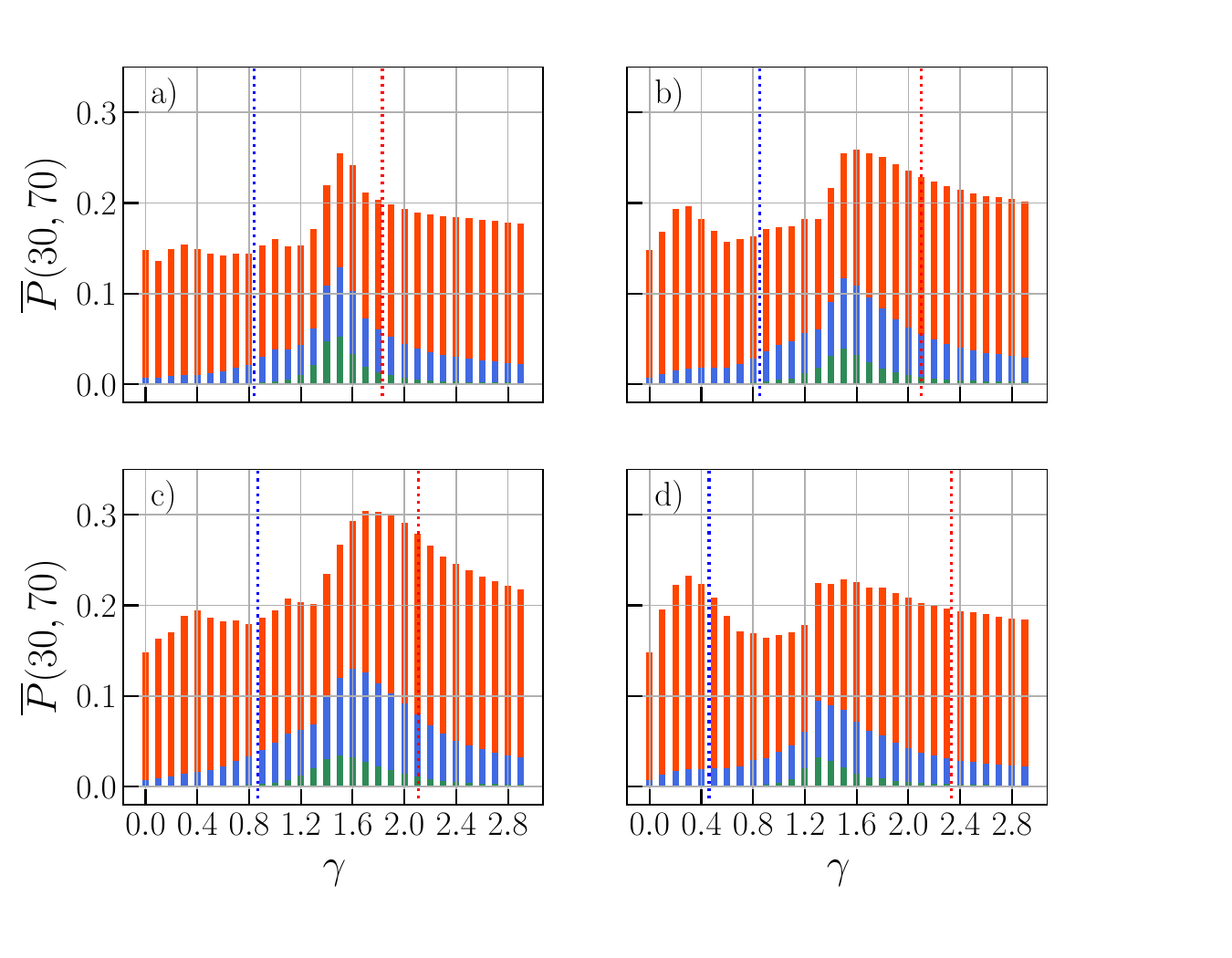}
    \caption{ Graph showing the success probability $\bar{P}(30, 70)$ versus $\gamma$ for the same (as figure \ref{fig:LHZ_plong_vs_gamma}) 5 qubit LHZ embedded SK spin glass instances decoded using `2+1' non-overlapping spanning trees. The total success probability at each $\gamma$ is split up into: `2+1' spanning trees correct (green), two spanning trees correct (blue) and one spanning tree correct (orange-red). The $\gamma_{\text{heur}}$ for the directly embedded instance and $\gamma_{\text{heur}}^{\text{LHZ}}$ are again plotted as dotted vertical lines in blue and red respectively.}
    \label{fig:LHZ_plong_split_vs_gamma}
\end{figure}

To test this theory, in figure \ref{fig:LHZ_plong_split_vs_gamma} we re-analysed the long-time success probability $\bar{P}(30, 70)$ versus $\gamma$ of the same four LHZ embedded 5 qubit SK spin glass instances decoded using the `2+1' non-overlapping spanning trees with lowest energy chosen, this time plotting $\bar{P}(30, 70)$ as bars split into where: three spanning trees were correct (green), two spanning trees were correct (blue) and when just one spanning tree was correct (orange-red).

We see that the fraction of bars that correspond to the success probability when all `2+1' or two spanning trees were correct, correlate with the success probabilities for the `2+1' non-overlapping spanning trees with majority vote. On the other hand, the fraction of the bars with only one spanning tree correct correlate with the improvement seen when using the non-overlapping and random spanning trees with lowest energy chosen to decode, in agreement with our theory.

We also note that the peak in success probability for one, two and `2+1' spanning trees correct is close to $\gamma^{\text{LHZ}}_{\text{heur}}$. However for the one spanning tree correct, clearly for instances b) and d), we also see a second peak in success probability closer to $\gamma_{\text{heur}}$ which is not clearly present in the two or `2+1' spanning tree correct fractions. It is an open question what the underlying cause of this second peak is, which we leave for future research.

\section{Success probability versus problem size: various decoding methods} \label{sec:prob_vs_size}

\begin{figure}
    \centering
    \includegraphics[width=0.45\textwidth]{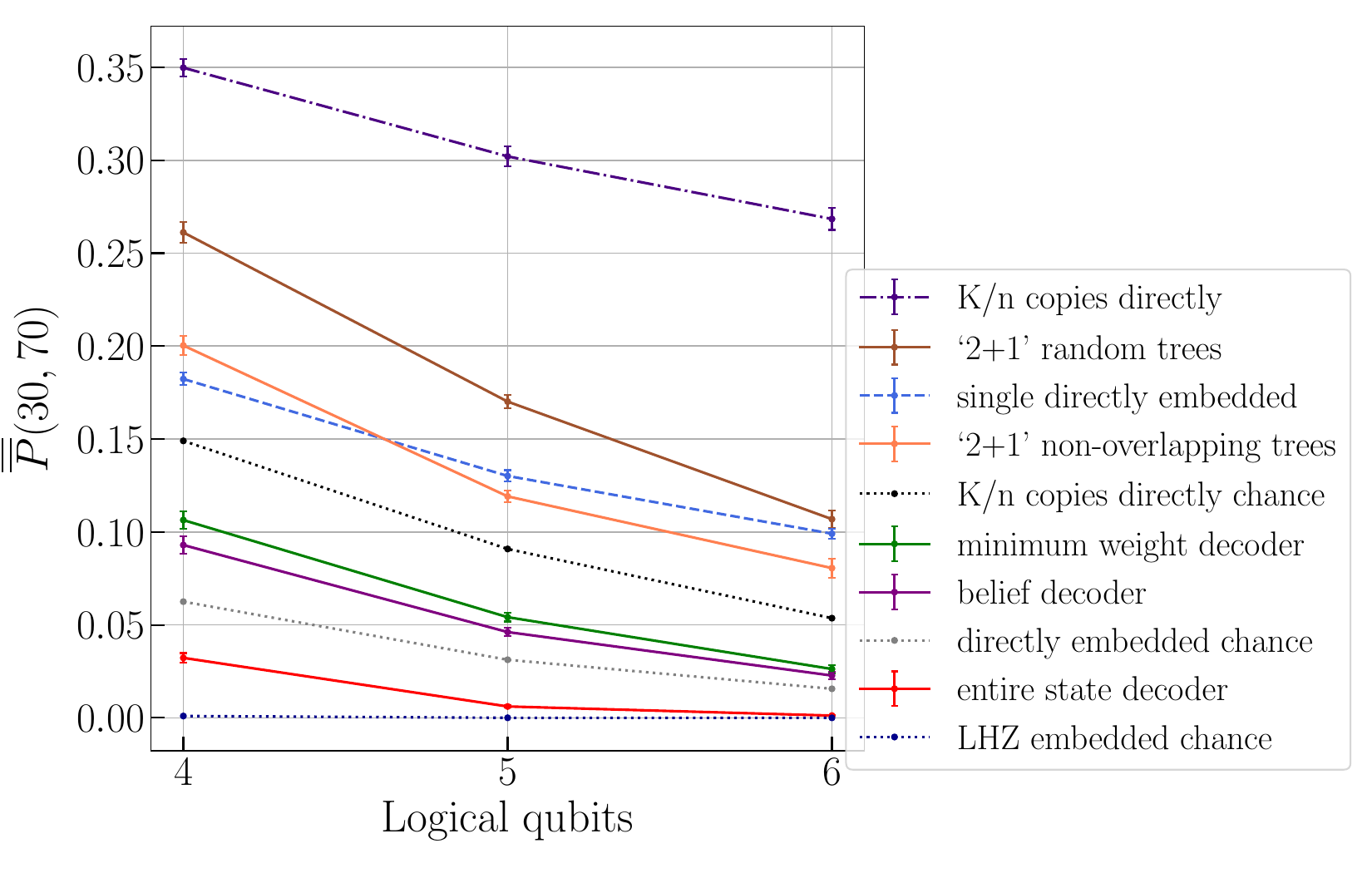}
    \caption{Graph showing $\overline{P}(30, 70)$ averaged over 100 instances of  4 and 5 logical qubit and 48 instances of 6 logical qubit, SK Ising spin glasses embedded onto the LHZ parity architecture versus the logical qubit number. The results are shown from five different decoding methods (solid lines): entire state (red), belief propagation (purple), minimum weight (green), `2+1' random spanning trees (salmon) and `2+1' non-overlapping spanning trees (brown). Error bars on each of the data points indicate the uncertainty in $\overline{\overline{P}}(30, 70)$. For comparison, the average success probability of $K/n$ copies and a single copy of the directly embedded instances are plotted in dot-dashed indigo and dashed blue respectively. The random chance probabilities (dotted) for: an LHZ embedded instance (dark blue), a single directly embedded instance (grey) and $K/n$ copies of a directly embedded instance (black) are also plotted.   }
    \label{fig:probs_vs_size}
\end{figure}

We next looked at how the performance of each of the decoding methods for the quantum walk scaled with the logical size of the LHZ parity embedded SK Ising spin glasses.
Figure \ref{fig:probs_vs_size} shows the averaged average long-time success probability $\overline{\overline{P}}(30, 70)$ of finding the ground state of LHZ parity embedded SK spin glass instances after decoding using several different methods versus the logical problem size of 4 (10 physical), 5 (15 physical) and 6 (21 physical) qubits. For 4 and 5 logical qubits the average was over 100 instances and for 6 logical qubits, 48 instances. The methods of decoding were: entire state decoding (solid red), belief decoder (solid purple), minimum weight decoder (solid green), `2+1' random spanning trees with lowest energy chosen (solid salmon) and `2+1' non-overlapping spanning trees with lowest energy chosen (solid brown). The `2+1' spanning trees signifies, that two spanning trees plus the set of data qubits were measured.

For comparison we also plotted the average success probability of $K/n$ copies (dot-dashed purple-indigo) and a single copy (dashed blue) of the directly embedded instances.
We also plotted the random chance success probability (before performing a quantum walk) of $K/n$ copies of the directly embedded model ($p_{K/n}$) (dotted black), the directly embedded model ($p_n=1/2^n$) (dotted gray) and the LHZ embedded model ($p_K=1/2^K$) (dotted dark blue). Note that $K/n$ is not an integer number of copies for 4 and 6 qubits, but corresponds to the same number of qubits as the LHZ embedded instance. We found $p_{K/n}$ using,
\begin{equation}
    p_{K/n} = 1 - (1 - p_n)^{K/n}.
\end{equation}

Looking at figure \ref{fig:probs_vs_size}, we see that all decoding methods outperform the random chance of finding the correct ground state on a LHZ embedded model. However, the LHZ embedded instances decoded using the entire state decoding method does not outperform the random chance of finding the correct ground state on the directly embedded model.
On the other hand, both the belief propagation and minimum weight decoders outperform the random chance of finding the correct ground state on the directly embedded model, but not the random chance of finding the correct ground state in $K/n$ copies of the directly embedded model.

Conversely, we see that both spanning tree decoding methods outperform the random chance of finding the correct ground state in $K/n$ copies of the directly embedded model. The `2+1' non-overlapping spanning trees outperform both the `2+1' random spanning trees and the single copies of directly embedded instances at the problem sizes shown. However, the scaling suggests that the single copies of directly embedded models would outperform both `2+1' spanning tree methods at larger sizes, though increasing the number of spanning trees measured could prevent this.
We also predict the advantage of the `2+1' non-overlapping trees over the `2+1' random trees would trend to zero as the size of the model increased due to the reducing likelihood of spanning trees overlapping. 

All LHZ parity embedded methods perform worse than $K/n$ copies directly embedded which use the same number of qubits (dot-dashed indigo). We note however the experimental impossibility of directly embedding all-to-all connected models particularly as problem size increases, meaning that some kind of embedding will have to be used. The results of this paper, indicate that by using LHZ parity embedding with the appropriate error correction techniques, we may be able to recover the success probability of the embedded model to a level close to what it would have been if a direct embedding of the model were possible.


\section{Conclusions and future directions}
\label{sec:conc_and_future}

In this work we have analysed and compared the performance of continuous-time quantum walks in solving LHZ parity embedded optimization problems, while using several different decoding methods.
We have found that the performance of the quantum walk on the LHZ parity embedded models approaches that of the `ideal' directly embedded models, when a decoding method which efficiently utilises the redundancy present in the LHZ parity architecture is selected.

When using the decoding methods: belief propagation, minimum weight and `2+1' non-overlapping spanning trees with majority vote, we found modest performance improvements. We suggest that this improvement is due to their ability to find correct logical states from correcting physical states which lie outside the logical codespace. We further suggest that their performance is limited, by the fact that their decoding selection process does not take into consideration the energy of potential choices of logical state. 

This suggestion is backed up by the fact that we see improved performance from the decoding methods: `2+1' random overlapping spanning trees and `2+1' non-overlapping spanning trees with lowest energy chosen. Both these methods take into account the energy of potential choices of logical state during their decoding selection process. This suggestion is further backed up by the fact that many of the correct ground states found by these techniques, are measured on just one spanning tree, something which would not be possible using a majority votes technique.

This finding suggests that other versions of decoders (such as an altered belief decoder) which take into account the energies of suggested states when selecting the final logical state, could also see enhanced improvement: this is an interesting future research direction. 
In addition, it suggests that error correction techniques for quantum computing in high noise scenarios should be tailored differently to those in low noise scenarios. 

We also found that, although the `2+1' non-overlapping spanning trees outperforms the single copies of the directly embedded instances for the logical problem sizes shown, its scaling is worse suggesting it would perform worse at larger sizes. However, in this decoding method, the number of logical state `choices' is fixed at three. It is an interesting open question if an alternative spanning tree method which increased the number of choices with logical problem size would perform better.

For the research in this paper, we compared the performance of LHZ embedded SK Ising spin glass models with different decoders against the performance of the equivalent (in number of physical qubits) number of copies of the same models but directly embedded. Although the directly embedded models are ideal and experimentally unlikely, their results provide a baseline to measure the success of the LHZ embedded results. 
As in an experimental situation it is extremely likely that some form of embedding would be needed, another interesting avenue of future research would be to compare the performance of alternative embeddings against each other.

We have also confirmed that the heuristic method developed in \cite{Callison2020} continues to be an effective way of estimating the value of the optimal hopping rate $\gamma_{\text{opt}}$. In addition, we confirmed that our results are in agreement with theory in \cite{Lanthaler2021} outlining the lower bound of the constraint strength $C$, if we take into account the increasing difference between the value of the optimal hopping rate $\gamma_{\text{opt}}$ and the heuristic hopping rate $\gamma_{\text{heur}}$ as the strength of $C$ increases.
We note that, due to the differences between the values of $\gamma_{\text{opt}}$ and $\gamma_{\text{heur}}$, the optimum of success probability for the instances with hopping rate of $\gamma_{\text{heur}}$ lies below the lower bound of $C$. However, to avoid any possible adverse affects caused by insufficient separation of the violating-constraint states with the non-violating-constraint states, for the research in this paper we fixed the value of $C=2$.
This raises the question, following the results in \cite{Nambu2024a, Nambu2024b, Nambu2025},
whether under error models where alternative decoding methods perform better, different constraint strengths than expected could be more beneficial. 

In the research presented here we have used the original (with the addition of data qubits) LHZ triangle parity embedding. In \cite{Ender2021}, the possibility of using alternative layouts, was first introduced. A interesting future direction might be to investigate the effect of these different layouts on the performance of quantum walks or other continuous-time quantum computing techniques.

Another open question is whether the improvement in performance accessed via the decoding methods for continuous-time quantum walks would also be present in hybrid forms of continuous-time quantum computing or in the case of quantum annealing on real hardware where there is significant presence of excited states. 
By varying anneal times, we could see what effect the number of excited states populated has on the performance of the decoding methods.

In addition, multi-stage quantum walks have shown promising performance in recent research \cite{Gerblich2024}, which potentially could be harnessed by utilising the embedding and decoding methods outlined in this research.
Alternate routes of future investigation could be looking at the effect of different graphs in the driver Hamiltonian such as the complete graph or others. Also the relation of the performance to the SK spin glasses' structure could be tested by using the random energy model \cite{Derrida1980} as was done in \cite{Callison2019}.

\section{Acknowledgements}
We thank Anita Weidinger for interesting discussions and Berend Klaver for providing error syndrome to belief propagation correction mappings.
This work was supported by the Austrian Research Promotion Agency (FFG Project No. FO999896208). This research was funded in whole, or in part, by the Austrian Science Fund (FWF) SFB BeyondC Project No. F7108-N38, through a START grant under Project No. Y1067-N27 and I 6011.
For the purpose of open access, the author has applied a CC BY public copyright licence to any Author Accepted Manuscript version arising from this submission.
This project was funded within the QuantERA II Programme that has received funding from the European Union's Horizon 2020 research and innovation programme under Grant Agreement No. 101017733.
 VK and NC were funded by UKRI EPSRC UK Quantum Technology Hub in Computing and Simulation (EP/T001062/1). We acknowledge funding from the the UKRI EPSRC International Network on Quantum Annealing (EP/W027003/1) was useful in the preparation of this work.
The computational results presented here have been achieved in part using the LEO HPC infrastructure of the University of Innsbruck.

\appendix

\section{The classical cost of measuring spanning trees}\label{apx:spannning_cost}

\bgroup
\def\arraystretch{1.5}
\begin{table}[]
\begin{tabular}{|l|l|l|l|l|}
\hline
m &\multicolumn{1}{|l|}{\begin{tabular}[c]{@{}l@{}}spanning tree\\ 01 02 03 12 13 23\end{tabular}} & new states   & \# new states   & total         \\ \hline
1 &\multicolumn{1}{|l|}{XXX \_ \_ \_}                                                              & XXX \_ \_ \_ & $2^3 = 8$ & 8 \\ \hline
2 &\multicolumn{1}{|l|}{XX\_ \_X\_}                                                                & XXY \_X\_    & $2^2 = 4$ & 12\\ \hline
3 &\multicolumn{1}{|l|}{XX\_ \_ \_ X}                                                              & XXY \_YX     & $2^1 = 2$ & 14 \\ \hline
4 &\multicolumn{1}{|l|}{X\_X X\_ \_}                                                               & XYX X\_ \_   & $2^2 = 4$ & 18 \\ \hline
5 &\multicolumn{1}{|l|}{X\_X \_ \_X}                                                               & XYX Y\_X     & $2^1 = 2$ & 20 \\ \hline
6 &\multicolumn{1}{|l|}{X\_ \_ XX\_}                                                               & XYY XX\_     & $2^1 = 2$ & 22 \\ \hline
7 &\multicolumn{1}{|l|}{X\_ \_ X\_X}                                                               & XYY XYX      & $2^0 = 1$ & 23 \\ \hline
8 &\multicolumn{1}{|l|}{X\_ \_ \_XX}                                                               & XYY YXX      & $2^0 = 1$ & 24 \\ \hline
9 & \multicolumn{1}{|l|}{\_XX X\_ \_}                                                               & YXX X \_ \_  & $2^2 = 4$ & 28 \\ \hline
10 &\multicolumn{1}{|l|}{\_XX \_X\_}                                                                & YXX YX\_     & $2^1 = 2$ & 30 \\ \hline
11 &\multicolumn{1}{|l|}{\_X\_ XX\_}                                                                & YXY XX\_     & $2^1 = 2$ & 32 \\ \hline
12 &\multicolumn{1}{|l|}{\_X\_ X\_X}                                                                & YXY XYX      & $2^0 = 1$ & 33 \\ \hline
13 &\multicolumn{1}{|l|}{\_X\_ \_XX}                                                                & YXY YXX      & $2^0 = 1$ & 34 \\ \hline
14 &\multicolumn{1}{|l|}{\_ \_X XX\_}                                                               & YYX XX\_     & $2^1 = 2$ & 36\\ \hline
15 &\multicolumn{1}{|l|}{\_ \_X X\_X}                                                              & YYX XYX      & $2^0 = 1$ & 37 \\ \hline
16 &\multicolumn{1}{|l|}{\_ \_X \_XX}                                                              & YYX YXX      & $2^0 = 1$ & 38 \\ \hline
                                           
\end{tabular}
\caption{Table showing the 16 possible spanning trees for the 4 logical qubit SK Ising spin glass on the 6 physical qubits of the parity encoded model in the second column. (The columns are counted left to right) The third column indicates the possible additional states (that haven't been counted yet), where `\_' indicates the qubit maybe in either 0 or 1 and `Y' indicates that it must be the opposite state to that which has appeared in a previous spanning tree (only one choice). The fourth column indicates the number of new states that the corresponding spanning tree has added and the fifth column shows the running total.}
\label{tab:new_trees_new_states}
\end{table}
\egroup

An instructive way to consider the cost of measuring spanning trees is to compare it to the cost of taking random guesses.

If we consider a 4 qubit logical Ising model problem, embedded onto an LHZ architecture with 6 qubits (here we exclude the contribution from the data qubits). Prior to performing a (quantum) computation and assuming a uniform probability distribution. If we decode using the entire state decoding method, it can be seen that we have a 1 in $2^6=64$ chance of getting the correct answer. The same chance as a random guess.
The effect of the quantum computation should be to at least increase the probability of measuring the correct ground state of the Ising model, even when using a simplistic decoding method such as entire state decoding. 

If we continue to look at the uniform probability distribution, and continue to decode using entire state decoding we can see that each time we take a new guess, the probability that we find the correct ground state increases with $g \times 1/64$, with $g$ being the number of guesses.

However if we instead used a random overlapping spanning trees method, we change how this probability behaves. If we just measure a single spanning tree, we already increase the number of physical states that we find which decode to the correct ground state.
This is because we only measure three qubits, for the spanning tree, meaning the other three are free to be in whichever state, meaning that there are 8 possible correct states out of the total of 64, meaning that we now have a 8 in 64 (or 1/8) chance of getting the correct answer.

If we were to measure one additional spanning tree, we might naively expect to add another $2^3 - 1 = 7$ possible states ($2^3$ minus the state where both spanning trees occur, which was already counted). Indeed this would be the case if we were to ensure that the additional spanning tree did not overlap with the original. In the uniform probability case this would mean that the chance of measuring the correct ground state when using two spanning trees is $(8+7)/64 = 15/64$.

However it is likely that if we were picking spanning trees (that are allowed to overlap) randomly, we would see an overlap of at least one qubit between spanning trees. This likelihood would increase with each additional spanning tree. (It is also impossible in the 4 logical, 6 physical qubit case for the third spanning tree in the two non-overlapping scenario, to not overlap.) When a spanning tree overlaps with another one, this means that the number of possible states that are added is reduced.
For instance, if the second spanning tree were to have overlapped with the original with just one qubit, then the number of additional possible states would have been $2^3 - 2^1 = 6$. If the overlap was two then it would have been $2^3-2^2 = 4$.
Generally we could say that number of additional states is $2^{(K-N)} - 2^{O_{AB}}$, where $O_{AB}$ is the overlap (in number of qubits) between the first two spanning trees $A, B$. If we were to add a third spanning tree $C$ then the additional number of states would be $2^{(K-N)} - 2^{O_{AC}} - (2^{O_{BC}}-2^{O_{ABC}})$. As the number of spanning trees increases, so does the number of ways they can overlap and so the number of terms in the above equations.

However, following a reasoning in table \ref{tab:new_trees_new_states}, we are able to count the growing number of additional states in a simpler way. First we write out (for K=6, n=4) the $n^{(n-2)} = 16$ possible spanning trees. To be general we denote the two possible qubit values as X (for correct as part of the spanning tree) and Y (for incorrect but not part of the current spanning tree). We denote qubits that may be in either state as `\_'. We then count the additional possible states by a process of elimination. 
For example the first tree XXX \_\_\_ adds $2^3$ possible states as there are three qubits that can be in either state. Then when we add the second tree XX\_ \_X\_, we notice that the states where the third qubit was X were already counted by the first spanning tree, so there is only one choice for the third qubit (Y), thus we have XXY \_X\_ so $2^2=4$ additional states. This happens again if we add the third spanning tree XX\_ \_\_X. Then we have XXY \_YX, so $2^1=2$ additional states.

If we continue this process of counting the additional states for each possible spanning tree (16 for 4 logical qubits) then we find that we end up with 38 possible states. In terms of probability, this represents for the uniform distribution, a 38/64 chance of finding the correct ground state, if we were to measure all the possible spanning trees. This shows the difference between the spanning tree decoding and repeated guesses of the entire state. If we were to continue guessing random entire states we would eventually find the ground state (success probability of 1). However even using every single spanning tree this approach will not find the ground state with certainty. 
This is because when measuring spanning trees, we are not altering the values of each of the physical qubits, but using them to decode from a physical to logical state.

We also note that a different order of spanning trees would add additional states at a different rate, i.e. two spanning trees with less overlap would add more states to the total. However, since there is a fixed total number of spanning trees, this means that the total number of states that can be added if all spanning trees are measured is fixed, even if the order in which the spanning trees are measured changes.


\section{Semi-analytical method for counting the total number of possible states introduced by measuring all spanning trees}\label{sec:find_sp_trees}

We start off by noting that a valid possible state, must have (in the X notation of the previous section), at least (n-1) X's.

If we look at the previous six qubit examples we can say that we can find the number of combinations of three X's  using the combination equation as follows,
\begin{equation}\label{eq:combinations}
    C = \frac{K!}{(K-n_X)!n_X!}.
\end{equation}
When $K=6$ and $n_X=3$, we find there are 20 possible combinations.

However we know that there exists a few invalid combinations i.e. those that do not correspond to spanning trees in the original model. Therefore we must subtract the number of invalid combinations from 20, in order to get the number of possible states with exactly three X's. We know that for this example, there are four invalid states: XX\_ X\_\_, X\_X \_X\_, \_XX \_\_X, and \_\_\_ XXX. Therefore there are 16 valid combinations  of three X's which correspond to possible states. As expected, this corresponds to the number of possible spanning trees on the four qubit SK spin glass. However we also need to know about the possible states with more than three X's.

If we next look at the number of combinations four X's, using equation \eqref{eq:combinations}, we can find that there are 15 possible combinations. In general we would normally need to now remove the combinations that correspond to invalid states, however in this example every combination containing four X's contains at least one valid spanning tree. Continuing the same steps for the permutations containing five and six qubits, we find that they add 6 and 1 possible state(s) respectively. If we sum all the valid possible states with three or more X's together, we find that this gives us a total of 38, which equals the total found using the method in table \ref{tab:new_trees_new_states}.

To describe this method more generally we could write the number of possible states $S$ as,
\begin{equation}\label{eq:poss_states}
    S = \sum_{i=l}^K \left( \frac{K!}{i!(K-i)!} - I_i\right),
\end{equation}
where $l$ is the length of the spanning trees $(n-1)$ and $I_i$ is the number of invalid states for that number of X's. 
When the number of $X$'s $n_X$ is equal to $l$, we are able to find $I_l$ by subtracting the number of possible spanning trees (found using Cayley's formula) from the number of combinations found using \eqref{eq:combinations}. However for $n_x > l$, how to find the number of invalid states $I_i$ remains an open question.

\section{Random overlapping spanning tree decoder: Performance vs number of trees} \label{sec:success_vs_n_trees}

\begin{figure}[!tb]
    \centering
    \includegraphics[width=0.45\textwidth]{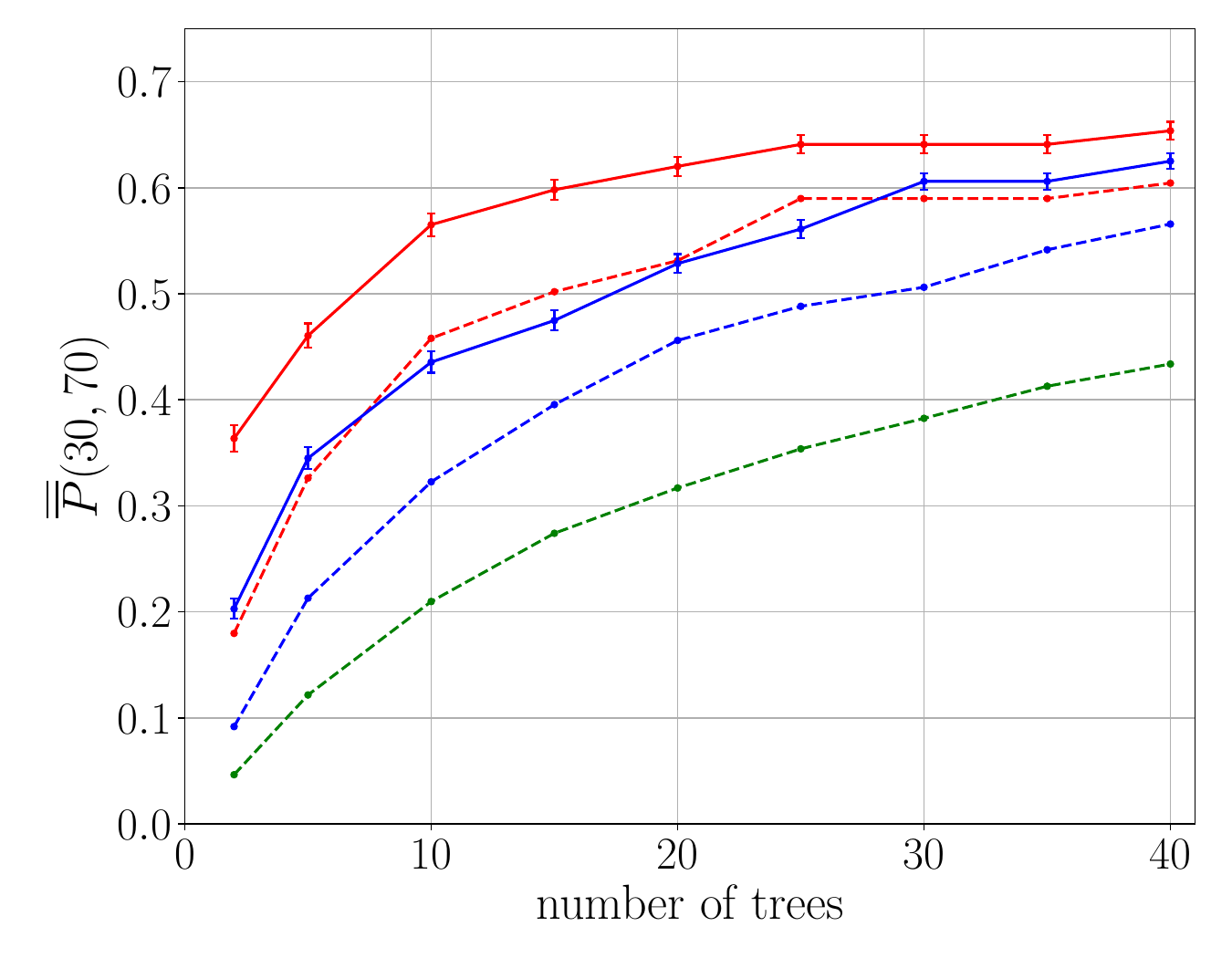}
    
    \caption{Graph showing the average success probability $\bar{P}(30, 70)$ averaged over 100 instances of 4 (red solid) and 5 (blue solid) logical qubit SK Ising spin glasses embedded onto the LHZ parity architecture, versus the number of random overlapping spanning trees. Error bars on each data point indicate the uncertainty in the averaged average success probability.  Plotted in the corresponding colour at each problem size, the dotted lines represent the number of states containing at least one correct spanning tree divided by the total number number of states ($2^K$), i.e. the probability of finding the correct ground state when decoding a randomly chosen state using random spanning trees. These data points do not have error bars as they are counted directly from the number of states determined by the order and number of spanning trees which is fixed at each problem size.}
    \label{fig:succ_prob_vs_span_trees}
\end{figure}

We next investigated the average performance when decoding using random overlapping spanning trees across 100 instances of LHZ parity embedded SK Ising spin glasses versus the number of random overlapping spanning trees measured for logical problem sizes of 4 (10 physical) and 5 (15 physical) qubits. 

Figure \ref{fig:succ_prob_vs_span_trees} shows the averaged average long-time success probability 
 $\overline{\overline{P}}(30, 70)$ vs number of spanning trees, over 100 instances of 4 (red), and 5 (blue) logical qubit LHZ parity embedded SK Ising spin glass instances. The standard error in $\overline{\overline{P}}(30, 70)$ is shown as error bars on each data point.
For comparison we also plotted the chance of finding the correct ground state from decoding a random physical state using random overlapping spanning trees versus the number of measured trees (dashed lines) for both 4 and 5 logical qubits in the corresponding colours and for 6 logical qubits in green.

We calculated this random chance by first finding the number of possible physical states containing at least one correct spanning tree at each number of spanning trees measured, and then dividing by the total number of possible physical states ($2^K$). We found the number of possible physical states containing at least one correct spanning tree using the method outlined in appendix \ref{sec:find_sp_trees}.

As the number of possible physical states which contain at least one correct spanning tree is determined solely by the number and position of the spanning trees, there are no error bars on these data points. We note however, that a different choice of spanning trees would change slightly the location and rate of increase of the data points. However as the total number of spanning trees is fixed, so is the total number of possible physical states containing a correct spanning tree, as we saw in appendix \ref{sec:find_sp_trees}. 

Looking at figure \ref{fig:succ_prob_vs_span_trees}, we see that as expected for both logical sizes, 4, and 5, the success probability of the quantum walk outperforms random chance. In both the quantum walk and random chance cases we see a reduction in rate of increase in the success probability as the number of spanning trees increase. This is due to the increased chance of a new spanning tree having an overlapping qubit with another spanning tree. An overlapping qubit causes a reduction in the number of new possible states and therefore the likelihood that a new correct spanning tree is found. We see no increase in success probability if a new spanning tree provides no new correct states or if a new spanning tree measures no new qubits (totally overlapping).

\bibliography{qw_LHZ.bib}

\end{document}